%% file: main.tex
\documentclass[print, aps, prd, nofootinbib,superscriptaddress]{revtex4-2}
\usepackage[T1]{fontenc}
\usepackage{amsmath,amssymb,graphicx, epsfig, bm, appendix}
\usepackage{natbib}
\usepackage{graphicx}
\usepackage[caption=false]{subfig}
\usepackage{xcolor, hyperref}
\usepackage{slashed}
\usepackage{comment}
\usepackage{booktabs}
\usepackage{orcidlink}
\usepackage[normalem]{ulem}

\input{macros}

\usepackage{makecell}

\hypersetup{colorlinks=true,		% false: colori base; true: personalizzazione colori
	linkcolor=blue(pigment),	          				% colore per link interni (immagini, tabelle, ...) (change box color with linkbordercolor)
	citecolor=blue(pigment),					% colore per riferimenti alla bibliografia
	filecolor=blue(pigment),							% colore per collegamenti a file
	urlcolor=blue(pigment)}							% colore per collegamenti a siti web

\allowdisplaybreaks 
\usepackage{float}
\usepackage[normalem]{ulem}
\usepackage{bbold}
\renewcommand{\d}{\mathrm{d}}

\def\nostrocostruttino#1\over#2{\mathrel{\mathop{\kern 0pt \rlap
{\hbox{$#1$}}} \hbox{\kern-.135em $#2$}}}

\usepackage{color}
\usepackage{cancel}
\begin{document} 

\title{Probing the Collins effect at photon-photon colliders}

\author{Simone Anedda \orcidlink{0009-0005-4921-2850}}
\email{sanedda@dsf.unica.it}
\affiliation{Dipartimento di Fisica, Universit\`a di Cagliari, Cittadella Universitaria, I-09042 Monserrato (CA), Italy}
\affiliation{INFN, Sezione di Cagliari, Cittadella Universitaria, I-09042 Monserrato (CA), Italy}

\author{Carlo Flore \orcidlink{0000-0002-1071-204X}}
\email{carlo.flore@unica.it}
\affiliation{Dipartimento di Fisica, Universit\`a di Cagliari, Cittadella Universitaria, I-09042 Monserrato (CA), Italy}
\affiliation{INFN, Sezione di Cagliari, Cittadella Universitaria, I-09042 Monserrato (CA), Italy}

\author{Francesco Murgia \orcidlink{0000-0003-0838-5518}}
\email{francesco.murgia@ca.infn.it}
\affiliation{INFN, Sezione di Cagliari, Cittadella Universitaria, I-09042 Monserrato (CA), Italy}

\author{Cristian Pisano \orcidlink{0000-0002-7717-1583}}
\email{cristian.pisano@unica.it}
\affiliation{Dipartimento di Fisica, Universit\`a di Cagliari, Cittadella Universitaria, I-09042 Monserrato (CA), Italy}
\affiliation{INFN, Sezione di Cagliari, Cittadella Universitaria, I-09042 Monserrato (CA), Italy}

\begin{abstract}
We propose to study the Collins effect in the production of almost back-to-back unpolarized hadrons (mainly charged pions) by looking at their azimuthal distributions around the jet thrust-axis in the process $\gamma\gamma \to h_1h_2X$.  The quasi-real photons in the initial state can  be emitted either by two scattering leptons or two charged heavy nuclei in ultra-peripheral collisions. We provide numerical estimates for kinematic configurations in principle accessible at the Future Circular Collider and at the planned experiments on ultra-peripheral collisions at the LHC. We find  that the corresponding asymmetries can be sizable, even up to around 10-15\%.  The proposed measurements would thus shed light on the fragmentation mechanism of quarks into hadrons and, in principle, would allow one to distinguish among different existing models for the Collins functions. 
\end{abstract}

\date{\today}
\maketitle
\input{Introduction}
\input{Formalism}

\input{Phenomenology}

\input{Conclusions}

\section*{Acknowledgments}
We thank C.~Cocuzza and D.~Pitonyak for providing us the JAM3D-2022 grids and for useful discussions. 
The work of S.A.\ and C.P.\ has been supported by Fondazione
di Sardegna through the project {\it Journey to the center of
the proton}, No.\ F23C25000150007.
C.F. is supported by the European Union’s Horizon Europe research and innovation programme under the Marie Sk{\lslash}odowska-Curie grant agreement n. 101150792 (STAT-TMDs).

\bibliographystyle{apsrev4-2}      % numerico, ordinato per citazione
\bibliography{references}            % file package.bib

\end{document}

%% file: macros.tex
\newcommand{\cf}[1]{{Fig.~\ref{#1}}}

\definecolor{persianindigo}{rgb}{0.2, 0.07, 0.48}
\definecolor{plum(traditional)}{rgb}{0.56, 0.27, 0.52}
\definecolor{purplemountainmajesty}{rgb}{0.59, 0.47, 0.71}
\definecolor{raspberryrose}{rgb}{0.7, 0.27, 0.42}
\definecolor{ruby}{rgb}{0.88, 0.07, 0.37}
\definecolor{bluebell}{rgb}{0.64, 0.64, 0.82}
\definecolor{ballblue}{rgb}{0.13, 0.67, 0.8}
\definecolor{blue(ncs)}{rgb}{0.0, 0.53, 0.74}
\definecolor{blue(pigment)}{rgb}{0.2, 0.2, 0.6}
\definecolor{forestgreen(web)}{rgb}{0.13, 0.55, 0.13}

\definecolor{pastelpink}{rgb}{0.75, 0.15, 0.49}

\newcommand{\lslash}{{\fontencoding{T1}\selectfont\symbol{170}}}

\def\d{{\rm d}}

\def\i-{\item[-]}

\def\1{\mathbb 1}

\def\avpperp{\langle p_\perp^2\rangle}
\def\gaga{\gamma\gamma}
\def\TOCA{TO-CA24 }
\def\JAM3D{JAM3D-2022-nolat}

%% file: Introduction.tex
\section{Introduction}
Hadronization, namely the mechanism by which the colored quarks and gluons produced in high-energy reactions turn into the colorless hadrons observed in the detectors, is one of the most challenging aspects of Quantum Chromodynamics (QCD). Being a nonperturbative process, it cannot be computed from first principles and, in the case of light hadrons, is described in terms of fragmentation functions (FFs).  Transverse momentum dependent (TMD) FFs provide more detailed information than the usual collinear FFs, 
as they depend not only 
on the  longitudinal momentum fraction $z$, but also on the intrinsic transverse momentum $\bm p_\perp$ of the final hadron with respect to the fragmenting parton. %Being nonperturbative objects, 
Their determination requires accurate phenomenological comparison with experimental measurements of observables for which TMD factorization has been proven, see e.g.~Ref.~\cite{Collins:2011zzd}. Concerning the extraction of quark TMD FFs, these observables include cross sections and azimuthal asymmetries in semi-inclusive deep inelastic scattering (SIDIS), $\ell p\, (N ) \to  \ell^\prime\, h\, X$, and  semi-inclusive $\ell^+\ell^-$  annihilation (SIA), that is the almost back-to-back production of two hadrons  in lepton-antilepton annihilations, $\ell^+ \ell^- \to \gamma^*, Z \to h_1\, h_2 \, X$. However, phenomenology in SIDIS is complicated by the fact that the FFs appear convoluted with TMD parton distribution functions, whereas for $e^+e^-$ SIA annihilation, where only fragmentation functions are involved, experimental data are presently relatively scarce. It would therefore be very important to consider other processes in order to improve our knowledge of TMD FFs.  

As recently pointed out in Ref.~\cite{Anedda:2025bts},  inclusive hadron-pair production in photon-photon collisions represent a class of reactions for which factorization is expected to hold because of the presence of only final state interactions, as in SIA, and could thus be used to probe TMD FFs. TMD factorization requires the presence in the process of two well distinct energy scales, $q_T$ and $Q$, such that $q_T \ll Q$. In these photon-photon processes the small scale is given by the transverse momentum of the two hadrons with respect to the thrust axis, which at leading order can be identified with the direction of the fragmenting quark-antiquark pair. On the other hand, the hard scale needed to perform the perturbative calculations for the underlying hard process $\gamma\gamma\to q\bar q$ can be identified (in the case of quasi-real photons, see below) with the transverse momentum of the jet thrust-axis with respect to the initial photon beams. An important advantage of this kind of probe would be a more efficient flavor separation than in SIDIS or $\ell^+\ell^-$ annihilations, since the contributions of $d, s, b$ quarks should be suppressed by a relative factor of 1/16 (rather than 1/4), coming from the involved fourth power of their fractional electric charge, with respect to those of $u, c$ quarks. This would help to disentangle the different flavor contributions to the quark TMD FFs. 

In contrast to Ref.~\cite{Anedda:2025bts} where, for a more formal direct comparison with the SIDIS process, the case of a deeply virtual and a quasi-real photon collision was considered, in this more phenomenological follow-up paper we focus only on processes initiated by quasi-real photons arising from two scattering leptons, or from two highly electrically charged heavy nuclei in ultra-peripheral collisions (UPCs). Moreover, we are mainly interested in the so-called Collins function~\cite{Collins:1992kk}, which describes the fragmentation of a transversely polarized quark into a non-collinear unpolarized hadron. As such, this function  generates azimuthal asymmetries in the distribution of the final hadrons around the jet-thrust axis. In order to assess the feasibility of the proposed measurements, we consider kinematic configurations that can in principle be studied at the Future Circular Collider (FCC-ee)  and in UPCs at the LHC. 

While several extractions of the unpolarized quark TMD FFs, entering the denominator of the Collins asymmetries, are available~\cite{Anselmino:2013lza,Bacchetta:2017gcc,Bacchetta:2022awv,Bacchetta:2024qre,Boglione:2022nzq,Scimemi:2019cmh,Moos:2025sal}, only a few extractions of the Collins FFs, entering the numerator of the asymmetry, have been performed \cite{Anselmino:2015sxa,Kang:2015msa,DAlesio:2020vtw, Gamberg:2022kdb,Boglione:2024dal}.
For our predictions, we will adopt the recent parametrizations in Refs.~\cite{Boglione:2024dal,Gamberg:2022kdb}.

The paper is organized as follows. In Section~\ref{sec:yy} we outline the main steps of the analytic calculation of the cross sections and the corresponding azimuthal asymmetries for the processes under study, assuming TMD factorization and adopting the helicity formalism. Furthermore, we present the explicit expression for the photon distributions used in our study and based on the Weizs\"acker-Williams (WW) approximation~\cite{vonWeizsacker:1934nji,Williams:1934ad}, first for unpolarized and longitudinally polarized photons inside leptons, and then for unpolarized photons inside heavy nuclei in UPCs.  Our numerical results for the FCC-ee kinematics and the UPC experiments planned at the LHC are presented in Section~\ref{sec:pheno}. Section~\ref{sec:conc} contains our summary and conclusions. 

%% file: Formalism.tex
\section{Hadron-pair production in two quasi-real photon collisions}\label{sec:yy}

In this section, adopting the same approach and formalism as in Ref.~\cite{Anedda:2025bts}, we will discuss the production of two unpolarized, almost back-to-back hadrons in quasi-real photon-photon collisions at lepton colliders and in heavy-ion UPCs. We will first focus on the case of lepton colliders, where the two quasi-real photons are emitted by the colliding leptons. At variance with Ref.~\cite{Anedda:2025bts}, here both final leptons are undetected and the photon flux is dominated by quasi-real, low-virtuality and almost collinear photons. The derivation of the hadron-pair production cross section and the asymmetries in the azimuthal distributions around the thrust-axis will follow the same lines as in the $\gamma^*\gamma$ case of Ref.~\cite{Anedda:2025bts}. Therefore, here we will restrict ourselves to the essential steps of the calculation. 
At the end of this section, we will show how the same results can be applied, with the appropriate modifications, to UPCs, where the quasi-real photons are coherently emitted by the two colliding heavy nuclei.

The process under consideration is the following:
\begin{equation}
\ell^+(l_+)\,\ell^-(l_-) \to\,h_1(P_1)\;h_2(P_2)\,+\,X\,,
\label{eq:A3:proc}
\end{equation}
through the photon-photon fusion channel
\begin{align}
\gamma_1\,(q_1)\;\gamma_2\,(q_2)\,
\to q(k_q)\;\bar{q}(k_{\bar q}) \,,
\end{align}
with $q_1^2 \approx q_2^2\approx 0$.
Assuming TMD factorization and within the helicity formalism, the invariant differential cross section for this process, at leading order in an electromagnetic and strong coupling constant power expansion, can be written as follows:
\begin{align}
 \d\sigma^{\ell^+\ell^-\to\, h_1\, h_2\, X} &=
 \frac{1}{4\,(l_+\cdot l_-)}\, \frac{\d^3\bm{k}_q}{2(2\pi)^3 k_q^{0}}\,\,\frac{\d^3\bm{k}_{\bar q}}{2(2\pi)^3 k_{\bar q}^{0}}\,(2\pi)^4\,\delta^{(4)}(q_1+q_2-k_q-k_{\bar q})\, \nonumber \\
 &\times \sum_q\,\sum_{\{\lambda_i\}}\,\rho_{\lambda_1,\lambda_1^\prime}(\gamma_1)\,\rho_{\lambda_2,\lambda_2^\prime}(\gamma_2)\,f_{\gamma_1/\ell^+,{\cal P}^{\ell^+}_{\hat z_+}}(\xi_1)\,\frac{\d\xi_1}{\xi_1}\,f_{\gamma_2/\ell^-,{\cal P}^{\ell^-}_{\hat z_-}}(\xi_2)\,\frac{\d\xi_2}{\xi_2}\nonumber\\ &\times \hat{H}_{\lambda_q,\lambda_{\bar{q}};\lambda_1,\lambda_2}\,\hat{H}^*_{\lambda^\prime_q,\lambda^\prime_{\bar{q}};\lambda_1^\prime,\lambda_2^\prime}\,
\hat{D}^{h_1}_{\lambda_q,\lambda^\prime_q}(z_1,\bm{p}_{\perp 1})\,\d z_1 \d^2\bm{p}_{\perp 1}\,
\hat{D}^{h_2}_{\lambda_{\bar{q}},\lambda^\prime_{\bar{q}}}\,(z_2,\bm{p}_{\perp 2})\,\d z_2 \d^2\bm{p}_{\perp 2}\,.
\label{eq:A3:gensig}
\end{align}
The first line includes all kinematical factors, the initial flux and the Lorentz invariant phase-space factors for the two observed jets, which in our leading order approach are identified with the final quark and antiquark in the hard partonic process. The second and third lines include all dynamical factors and will be denoted (up to the factor ${\rm d}\xi_1{\rm d}\xi_2/(\xi_1\xi_2)$\,) as the Kernel of Eq.~\eqref{eq:A3:gensig} in the sequel. See Ref.~\cite{Anedda:2025bts} for further details.

In the helicity formalism, the dynamical contribution is divided into three steps describing respectively: the polarization state of the two initial photons; the polarization transfer in the hard partonic scattering $\gamma\gamma\to q\bar q$; the (polarized) hadronization of the two final partons into the two observed hadrons inside the final jets.

In the leptonic center of mass (cm) reference frame and neglecting all masses, we have:
\begin{align}
l_+\, =\, \frac{\sqrt{s}}{2}&\,(\,1,0,0,1\,),\qquad l_-\, =\, \frac{\sqrt{s}}{2}\,(\,1,0,0,-1\,)\,,\qquad
q_1\,=\, \xi_1\,\frac{\sqrt{s}}{2}\,(\,1,0,0,1\,),\qquad q_2\, =\, \xi_2\,\frac{\sqrt{s}}{2}\,(\,1,0,0,-1\,)\nonumber\\
&q_1+q_2\, =\, \frac{\sqrt{s}}{2}\,(\xi_1+\xi_2,0,0,\xi_1-\xi_2\,),\qquad
k_{q,\bar{q}}\,=\,k_T\,(\,\cosh\eta_{q,\bar q},\pm \cos\phi_q, \pm \sin\phi_q,\sinh\eta_{q,\bar q}\,)\,,
\label{eq:A3:kin-lep}
\end{align}
where $\xi_{1,2}$ are the light-cone momentum fractions of photons 1 and 2 with respect to the parent leptons; $\eta_{q,\bar{q}}$ are the rapidities of the quark and antiquark respectively, in the leptonic cm frame; $k_T\,=\,|\bm{k}_T|$, with $\bm{k}_T\,=\,k_T\,(\,\cos\phi_q,\sin\phi_q\,)$\, is the quark (jet) transverse momentum;
$\phi_q$ is the azimuthal angle of the plane containing the thrust and the lepton-beam axes with respect to the $xz$-plane of the chosen leptonic cm frame.

From Eq.~\eqref{eq:A3:kin-lep} we see that the total energy squared in the partonic cm reference frame is $\hat s\,=\,\xi_1\,\xi_2\,s$.
Moreover, in this frame,
\begin{align}
q_1\,&=\,\frac{\sqrt{\hat s}}{2}\,(\,1,0,0,1\,),\qquad q_2\, =\, \frac{\sqrt{\hat s}}{2}\,(\,1,0,0,-1\,)\nonumber\\
k_{q,\bar{q}}\,&=\,k_T(\,\cosh\hat{\eta},\pm \cos\phi_q, \pm \sin\phi_q,\pm \sinh\hat{\eta}\,)\,,
\label{eq:A3:kin-par}
\end{align}
where $\hat\eta = \hat\eta_q  = -\hat\eta_{\bar q} = (\,\eta_q - \eta_{\bar q}\,)/2$\,.

Let us now consider the kinematic factors in the first line of Eq.~\eqref{eq:A3:gensig}. The flux factor is $4\,(l_+\cdot l_-)\,=\,2\,s$.
Working in the leptonic cm frame, the four-dimensional Dirac delta can be easily decomposed as follows:
\begin{align}
&\,\delta^{(4)}(q_1+q_2-k_q-k_{\bar q})\,=
\,\frac{2}{s}\,\delta\Bigl(\,\xi_1\,-\,\frac{k_T}{\sqrt{s}}\,\bigl(\,e^{\eta_q}\,+\,e^{\eta_{\bar q}}\bigr)\,\Bigr)\,\nonumber\\
&\times\,\delta\Bigl(\,\xi_2\,-\,\frac{k_T}{\sqrt{s}}\,\bigl(\,e^{-\eta_q}\,+\,e^{-\eta_{\bar q}}\bigr)\,\Bigr)\,\delta^{(2)}\bigl(\,-\bm{k}_{q\,T}-\bm{k}_{\bar{q}\,T} \,\bigr)\,.
\label{eq:A3:delta4}
\end{align}

Therefore, the light-cone momentum fractions $\xi_{1,2}$ of the quasi-real photons are fixed once we know the transverse momentum $k_T$ and the rapidities $\eta_{q,\bar{q}}$ of the two jets originating from the $q,\bar{q}$ fragmentation. More precisely,
\begin{align}
\bar{\xi}_1\,=\,\frac{k_T}{\sqrt{s}}\,\bigl(\,e^{\eta_q}\,+\,e^{\eta_{\bar q}}\bigr)\,,\qquad
\bar{\xi}_2\,=\,\frac{k_T}{\sqrt{s}}\,\bigl(\,e^{-\eta_q}\,+\,e^{-\eta_{\bar q}}\bigr)\,.
\label{eq:A3:csi12-fix}
\end{align}

Clearly, the remaining two-dimensional Dirac delta in Eq.~\eqref{eq:A3:delta4} gives $\bm{k}_{q\,T}\equiv\bm{k}_T = -\bm{k}_{\bar{q}\,T}$, as anticipated in Eqs.~\eqref{eq:A3:kin-lep} and~\eqref{eq:A3:kin-par}.

Moreover, we can write
\begin{equation}\label{eq:A3:d3qqbar}
\frac{\d^3\bm{k}_q}{2(2\pi)^3\, k_q^{0}}\,\frac{\d^3\bm{k}_{\bar q}}{2(2\pi)^3 k_{\bar q}^{0}}\,\delta^{(2)}\bigl(\,-\bm{k}_{q\,T}-\bm{k}_{\bar{q}\,T} \,\bigr)\,
=\,\frac{1}{2^8\,\pi^6}\,\d^2\bm{k}_T\,\d\eta_q\,\d\eta_{\bar q}\,,
\end{equation}
where we have used Eq.~(\ref{eq:A3:kin-par}) to set $\d k_{q,\bar q}^3/k_{q,\bar q}^0\,=\,\d \eta_{q,\bar q}$\,.

The expression of the differential cross section in Eq.~\eqref{eq:A3:gensig} can therefore be rewritten as:
\begin{align}
\d \sigma^{\ell^+\ell^-\,\to\,h_1\,h_2\,X} &=\,\frac{1}{2\,s}\,\frac{2}{s}\,\delta\Bigl(\,\xi_1\,-\,\frac{k_T}{\sqrt{s}}\,\bigl(\,e^{\eta_q}\,+\,e^{\eta_{\bar q}}\bigr)\,\Bigr)\,\delta\Bigl(\,\xi_2\,-\,\frac{k_T}{\sqrt{s}}\,\bigl(\,e^{-\eta_q}\,+\,e^{-\eta_{\bar q}}\bigr)\,\Bigr)\nonumber\\
&\times\,(2\pi)^4\,\frac{1}{2^8\,\pi^6}\,\d^2\bm{k}_T\,\d\eta_q\,\d\eta_{\bar q}\,\frac{\d \xi_1}{\xi_1}\,\frac{\d \xi_2}{\xi_2}\,\times\,{\rm Kernel}\nonumber\\
&=\frac{1}{16\,\pi^2\,s^2}\,\frac{1}{\bar{\xi}_1\,\bar{\xi}_2}\,\d^2\bm{k}_T\,\d\eta_q\,\d\eta_{\bar q}\,\times\,{\rm Kernel}\,.
\label{eq:A3:gensig-2}
\end{align}

Furthermore, from Eq.~\eqref{eq:A3:csi12-fix},
\begin{equation}
\frac{1}{\bar{\xi}_1\,\bar{\xi}_2}\,=\,\frac{s}{k_T^2}\,\frac{1}{2\,[\,1+\cosh(\eta_q\,-\,\eta_{\bar q})\,]}\,,
\label{eq:A3:xi-cosh}
\end{equation}
so that
\begin{align}
\frac{\d \sigma^{\ell^+\ell^-\,\to\,h_1\,h_2\,X}}{\d^2\bm{k}_T\,\d\eta_q\,\d\eta_{\bar q}}\,=\,\frac{1}{32\,\pi^2\,s\,k_T^2}\,\frac{1}{1+\cosh(\eta_q\,-\,\eta_{\bar q})}\,\times\,{\rm Kernel}\,.
\label{eq:dsig-full.1}
\end{align}

Let us now consider all the factors in the dynamical Kernel, as defined in the last two lines of Eq.~\eqref{eq:A3:gensig}. Since they were already discussed in detail in Ref.~\cite{Anedda:2025bts}, we will only briefly recall the main steps of the calculation here.

The normalized helicity density matrix of the two quasi-real photons reduces to
\begin{equation}
\rho_{\lambda, \lambda^\prime}(\gamma_i) = \frac{1}{2}\,
\begin{pmatrix}
1 + {\cal P}^{\gamma_i}_{\hat z_i} & 0 \\
0 & 1 - {\cal P}^{\gamma_i}_{\hat z_i}
\end{pmatrix}\,,
\label{eq:hel-density-matrix}
\end{equation}
where ${\cal P}^{\gamma_i}_{\hat z_i}$
is the longitudinal component of the polarization vector of the photon $i$ in its own helicity reference frame.
Due to parity conservation, only two combinations of the photon WW distribution inside the initial leptons may appear in the Kernel:
\begin{align}
&\big[\,\rho_{++}(\gamma_i)\,+\,\rho_{--}(\gamma_i)\,\bigr]\,f_{\gamma_i/\ell^{\pm},{\cal P}_{\pm}}(\bar{\xi}_i)\nonumber\\
&\,=\,f_{\gamma_i,+/\ell^{\pm},{\cal P}_{\pm}}\,(\bar{\xi}_i)\,+\,f_{\gamma_i,-/\ell^{\pm},{\cal P}_{\pm}}\,(\bar{\xi}_i)\,\equiv\,f_{\gamma_i/\ell^{\pm}}(\bar{\xi}_i)\,,
\label{eq:A3:ww-unp}
\end{align}
and
\begin{align}
&\big[\,\rho_{++}(\gamma_i)\,-\,\rho_{--}(\gamma_i)\,\bigr]\,f_{\gamma_i/\ell^{\pm},{\cal P}_{\pm}}(\bar{\xi}_i)\nonumber\\
&\,=\,f_{\gamma_i,+/\ell^{\pm},{\cal P}_{\pm}}\,(\bar{\xi}_i)\,-\,f_{\gamma_i,-/\ell^{\pm},{\cal P}_{\pm}}\,(\bar{\xi}_i)\,\equiv
{\cal P}_{\pm}\,\Delta_L f_{\gamma_i/\ell^{\pm}}(\bar{\xi}_i)\,,
\label{eq:A3:ww-pol}
\end{align}
where we use the shorthand notation $ {\cal P}_{\pm}\,\equiv\,{\cal P}^{\ell^{\pm}}_{\hat z_{\pm}}$ for the longitudinal component of the lepton polarization vectors (in their own helicity reference frame), with ${\cal P}_{\pm}\,=\,0,\,\pm 1$.

Concerning the helicity amplitudes $\hat{H}_{\lambda_q,\lambda_{\bar{q}};\lambda_1,\lambda_2}$ for the partonic hard scattering process,
\begin{equation}
\gamma_1(q_1,\lambda_1)\,\gamma_2(q_2,\lambda_2)\,\to\,q(k_q,\lambda_q)\,\bar{q}(k_{\bar q},\lambda_{\bar q})\,,
\label{eq:A3:hard-proc}
\end{equation}
they can be directly recovered from the ones given in Eq.~(C1) of Ref.~\cite{Anedda:2025bts} for the $\gamma^*\,\gamma$ case, by simply taking the limit $Q^2\,=\,- q_1^2\,\to\,0$. As a result, only four amplitudes are non-vanishing (in the massless limit considered here) and due to parity conservation only two of them are independent:
\begin{align}
\hat{H}_{+-;1,-1}\,&=\,-\,\hat{H}^*_{-+;-1,1}\,=\,-\,2\,\sqrt{3}\,e^2\,e_q^2\,e^{i2\phi_q}\,e^{(\eta_q-\eta_{\bar q})/2}\,,\nonumber\\
\hat{H}_{+-;-1,1}\,&=\,-\,\hat{H}^*_{-+;1,-1}\,=\,2\,\sqrt{3}\,e^2\,e_q^2\,e^{-i2\phi_q}\,e^{-(\eta_q-\eta_{\bar q})/2}\,.
\label{eq:A3:Hel}
\end{align}
Regarding the fragmentation sector, for (pseudo)scalar or unpolarized hadrons only two terms contribute to the kernel, involving respectively the product of the unpolarized fragmentation functions,%\simo{[sempre in base all' intro, (FFs)]}
$\,D_{h_1/q}(z_1,p_{\perp 1})\,D_{h_2/\bar q}(z_2,p_{\perp 2})$, and that of the Collins FFs, $\,\Delta^N D_{h_1/q^\uparrow}(z_1,\bm{p}_{\perp 1})\,\Delta^N D_{h_2/\bar{q}^\uparrow}(z_2,\bm{p}_{\perp 2})$. See also Section II and Eqs.~(17)-(23) of Ref.~\cite{Anedda:2025bts} for more details.

Inserting all these ingredients into Eq.~\eqref{eq:dsig-full.1}, after some calculations, we finally get:
\begin{align}
&\frac{\d \sigma^{\ell^+\ell^-\,\to\,h_1\,h_2\,X}({\cal P}_+,{\cal P}_-)}{\d^2\bm{k}_T\,\d\eta_q\,\d\eta_{\bar q}\,\d z_1\,\d^2\bm{p}_{\perp 1}\,\d z_2\,\d^2\bm{p}_{\perp 2}}\,=\,\frac{6\,\alpha^2}{s\,k_T^2}\,\frac{1}{1+\cosh(\eta_q\,-\,\eta_{\bar q})}\,\sum_q e_q^4\nonumber\\
&\qquad \times \bigl\{\,f_{\gamma/\ell}(\bar{\xi}_1)\,f_{\gamma/\ell}(\bar{\xi}_2)\,-\,{\cal P}_+\,{\cal P}_-\,\Delta_L f_{\gamma/\ell}(\bar{\xi}_1)\,
\Delta_L f_{\gamma/\ell}(\bar{\xi}_2)\,\bigr\}\nonumber\\
&\qquad \times \Bigl\{\,\cosh(\eta_q\,-\,\eta_{\bar q})\,D_{h_1/q}(z_1,p_{\perp 1})\,D_{h_2/\bar q}(z_2,p_{\perp 2})\,\nonumber\\
&\qquad +\,\frac{1}{4}\,\cos(\phi_q^{h_1}\,-\,\phi_{\bar q}^{h_2})\,\Delta^N D_{h_1/q^\uparrow}(z_1,p_{\perp 1})\,\Delta^N D_{h_2/\bar{q}^\uparrow}(z_2,p_{\perp 2})\,\Bigr\}\,,
\label{eq:dsig-full}
\end{align}
where $\phi_q^{h_1}$ and $\phi_{\bar q}^{h_2}$ are respectively the azimuthal angles of the two hadron momenta around the direction of motion of the parent quark/antiquark, as measured in the parton helicity reference frames.
Notice that, at variance with the more complex $\gamma^*\gamma$ case, here the unpolarized and longitudinally polarized WW distributions multiply the same factors related to the fragmentation functions; therefore, their contributions are simply additive. By introducing the variables
\begin{equation}
\zeta_q\,\equiv\,\zeta\,=\,\frac{1}{1+e^{\eta_q-\eta_{\bar q}}}\,\qquad\,\zeta_{\bar q}\,=\,1-\zeta\,=\,\frac{1}{1+e^{-(\eta_q-\eta_{\bar q})}}\,,
\label{eq:A3:zeta}
\end{equation}
so that $0 \leq \zeta \leq 1$,
we find that
\begin{align}
\frac{\cosh(\eta_q-\eta_{\bar q})}{1+\cosh(\eta_q-\eta_{\bar q})} &=\,\zeta^2\,+\,(1-\zeta)^2\,,\nonumber\\
\frac{1}{1+\cosh(\eta_q-\eta_{\bar q})} &=\,2\,\zeta\,(1-\zeta)\,.
\label{eq:A3:etazeta}
\end{align}

It is therefore easy to see that Eq.~\eqref{eq:dsig-full} agrees, with obvious changes in the notation, with Eq.~(2.93) of Ref.~\cite{Anedda:2026xeo}, derived independently for the case of heavy-ion UPCs. Notice that, there, the azimuthal angles $\phi_{1,2}$ are both measured around the chosen jet light-cone direction, while in our case, as said, $\phi_q^{h_1},\,\phi_{\bar q}^{h_2}$ are each measured around the $\hat z$ axis of the parent parton helicity frame. Therefore, $\phi_q^{h_1}\,=\,\phi_1$, $\phi_{\bar q}^{h_2}\,=\,2\,\pi\,-\,\phi_2$\,.

Since $k_T$ plays the role of the hard scale for this process, to guarantee the validity of the TMD approach we require that $k_T \geq k_0 \gg \Lambda_{\rm QCD}$, where $k_0$ is a convenient reference hard scale. Therefore, the allowed range of values for $k_T$ is
\begin{equation}
k_0\, \leq\, k_T\, \leq \, \frac{\sqrt{s}}{2}\,, 
\label{eq:A3:ktrange}
\end{equation}
with $k_T = \sqrt{s}/{2}$ corresponding to the kinematical  limit $\eta_q = \eta_{\bar q} = 0$ and $\bar{\xi}_1 = \bar{\xi}_2 = 1$. Furthermore, one can see that at fixed $s$ and $k_T$, it must be
\begin{equation}
\log\Biggl\{\frac{1}{2}\,\Bigl[\,\frac{\sqrt{s}}{k_T}-\sqrt{\frac{s}{k_T^2}-4}\,\Bigr]\,\Biggr\}\,\leq\,\eta_q\,\leq\,\log\Biggl\{\frac{1}{2}\,\Bigl[\,\frac{\sqrt{s}}{k_T}+\sqrt{\frac{s}{k_T^2}-4}\,\Bigr]\,\Biggr\}\,.
\label{eq:A3:etaq-lim}
\end{equation}
Notice that $\sqrt{s}/k_T\,\geq\,2$,  because of Eq.~\eqref{eq:A3:ktrange}. Finally, at fixed $\eta_q$ within the range in Eq~.\eqref{eq:A3:etaq-lim}, one has
\begin{equation}
-\,\log\Bigl(\frac{\sqrt{s}}{k_T}\,-\,e^{-\eta_q}\,\Bigr)\,\leq\,\eta_{\bar q}\,\leq\,\log\Bigl(\frac{\sqrt{s}}{k_T}\,-\,e^{\eta_q}\,\Bigr)\,.
\label{eq:A3:etab-lim}
\end{equation}
We point out that the allowed phase space for $\eta_q$ and $\eta_{\bar q}$ is clearly symmetric.

To collect more statistics, we can first consider that
\[\d^2\bm{k}_T = k_T\,\d k_T\,\d\phi_q = (1/2)\,\d k_T^2\,\d\phi_q\,.\]
Using the fact that the differential cross section, Eq.~\eqref{eq:dsig-full}, is independent of $\phi_q$, we can integrate over it and simply get an additional $2\pi$ factor on the right-hand side. Moreover, using Eqs.~(22) and (23) of Ref.~\cite{Anedda:2025bts}, and performing the substitution $(\phi_q^{h_1},\phi_{\bar q}^{h_2})\,\to\,(\phi_q^{h_1},\phi_q^{h_1}-\phi_{\bar q}^{h_2}\,\equiv\,\phi_{12})$ we can write
\begin{align}
&\int \d^2\bm{p}_{\perp 1}\int \d^2\bm{p}_{\perp 2}\,D_{h_1/q}(z_1,p_{\perp 1})\,D_{h_2/\bar q}(z_2,p_{\perp 2})\,=\,D_{h_1/q}(z_1)\,D_{h_2/\bar q}(z_2)\,,\nonumber\\
&\int \d^2\bm{p}_{\perp 1}\int \d^2\bm{p}_{\perp 2}\,\Delta^N D_{h_1/q^\uparrow}(z_1,p_{\perp 1})\,\Delta^N D_{h_2/{\bar q}^\uparrow}(z_2,p_{\perp 2})\nonumber\\
&\,\qquad\qquad\quad=\,\frac{1}{2\pi}\,\Delta^N D_{h_1/q^\uparrow}(z_1)\,\Delta^N D_{h_2/{\bar q}^\uparrow}(z_2)\,\d\phi_{12}\,.
\label{eq:A3:FFmom}
\end{align}

Inserting these results into Eq.~\eqref{eq:dsig-full} we therefore get the less differential cross section
\begin{align}
&\frac{\d \sigma^{\ell^+\ell^-\,\to\,h_1\,h_2\,X}({\cal P}_+,{\cal P}_-)}{\d {k}_T^2\,\d\eta_q\,\d\eta_{\bar q}\,\d z_1\,\d z_2\,\d\phi_{12}}\,=\,\frac{3\,\alpha^2}{s\,{k}_T^2}\,\frac{1}{1+\cosh(\eta_q\,-\,\eta_{\bar q})}\,\sum_q e_q^4\nonumber\\
&\qquad \times \bigl\{\,f_{\gamma/\ell}(\bar{\xi}_1)\,f_{\gamma/\ell}(\bar{\xi}_2)\,-\,{\cal P}_+\,{\cal P}_-\,\Delta_L f_{\gamma/\ell}(\bar{\xi}_1)\,
\Delta_L f_{\gamma/\ell}(\bar{\xi}_2)\,\bigr\}\nonumber\\
&\qquad \times \Bigl\{\,\cosh(\eta_q\,-\,\eta_{\bar q})\,D_{h_1/q}(z_1)\,D_{h_2/\bar q}(z_2)\,\nonumber\\
&\qquad +\,\frac{1}{4}\,\cos\phi_{12}\,\Delta^N D_{h_1/q^\uparrow}(z_1)\,\Delta^N D_{h_2/\bar{q}^\uparrow}(z_2)\,\Bigr\}.
\label{eq:dsig-pt-int}
\end{align}

Adopting the shorthand notation $\d\sigma({\cal P}_+,{\cal P}_-)$ for the differential cross section, we can write
\begin{align}
\d\sigma^{\rm unp}\,&=\,\frac{1}{2}\,\bigl[\,\d\sigma(+,+)\,+\,\d\sigma(+,-)\bigr]\,,\nonumber\\
\d\Delta_L\sigma\,&=\,\d\sigma(+,+)\,-\,\d\sigma(+,-)\,,
\label{eq:A3:ds-ll}
\end{align}
and define the double longitudinal spin asymmetry
\begin{align}
A_{LL}\,&=\,\frac{\d\sigma(+,+)\,-\,\d\sigma(+,-)}{\d\sigma(+,+)\,+\,\d\sigma(+,-)}\,=\,\frac{\d\Delta_L\sigma}{2\,\d\sigma^{\rm unp}}\,.
\label{eq:A3:all}
\end{align}

We can also define azimuthal moments of the differential cross section $\d\sigma^{\rm unp}$ and the double longitudinal spin asymmetry $A_{LL}$:

\begin{align}
\langle\,\d\sigma^{\rm unp}\,|\,\cos\phi_{12}\,\rangle\,&=\,2\,\frac{\int \d\phi_{12}\,\cos\phi_{12}\,\d\sigma^{\rm unp}}{\int \d\phi_{12}\,\d\sigma^{\rm unp}}\nonumber\\
&=\,\frac{1}{4\,\cosh(\eta_q\,-\,\eta_{\bar q})}\,\frac{\sum_q\,e_q^4\,\Delta^N D_{h_1/q^\uparrow}(z_1)\,\Delta^N D_{h_2/\bar{q}^\uparrow}(z_2)}{\sum_q\,e_q^4\,D_{h_1/q}(z_1)\,D_{h_2/\bar q}(z_2)}\,,\nonumber\\ \nonumber\\
\langle\,A_{LL}\,|\,\cos\phi_{12}\,\rangle\,&=\,2\,\frac{\int \d\phi_{12}\,\cos\phi_{12}\,A_{LL}\,\d\sigma^{\rm unp}}{\int \d\phi_{12}\,\d\sigma^{\rm unp}}\nonumber\\
&\mkern-40mu =\,-\,\frac{1}{4\,\cosh(\eta_q\,-\,\eta_{\bar q})}\,\frac{\Delta_L f_{\gamma/\ell}(\bar\xi_1)\,\Delta_L f_{\gamma/\ell}(\bar\xi_2)}{f_{\gamma/\ell}(\bar\xi_1)\,f_{\gamma/\ell}(\bar\xi_2)}\,\frac{\sum_q\,e_q^4\,\Delta^N D_{h_1/q^\uparrow}(z_1)\,\Delta^N D_{h_2/\bar{q}^\uparrow}(z_2)}{\sum_q\,e_q^4\,D_{h_1/q}(z_1)\,D_{h_2/\bar q}(z_2)}\,,
\label{eq:azimom}
\end{align}
that still depend on $k_T$, $\eta_q$, $\eta_{\bar q}$, $z_1$ and $z_2$.

These results are obtained by simplifying all common factors within the numerator and denominator of the azimuthal moments. Clearly this procedure is only justified if we are working at fixed kinematical variables $k_T$, $\eta_{q,\bar q}\,$. In case we want to integrate one or more of these variables over some definite range we need to reinstate all appropriate factors in the numerator and denominator of the moments. This is because $\bar\xi_{1,2}$ in Eq.~\eqref{eq:A3:csi12-fix} explicitly depend on $k_T$ and $\eta_{q,\bar q}\,$. Additionally, both the WW photon distributions and the $q$, $\bar q$ fragmentation functions depend, through scale evolution (which is always understood here), on the hard scale of the process, $k_T\,$.

Let us also remark that the above discussion on the azimuthal asymmetries and moments, based on Eq.~\eqref{eq:dsig-pt-int}, applies as well to the more differential cross section of Eq.~\eqref{eq:dsig-full}, where the intrinsic transverse momenta of the two hadrons have not been integrated out.

Finally, let us comment on the case of two heavy-nuclei ultra-peripheral collisions,
\begin{equation}
{\cal N}_1({\cal P}_1)\,{\cal N}_2({\cal P}_2)\,\to \,h_1(P_1)\, h_2(P_2)\,+\,X\,,
\label{eq:UPC-proc}
\end{equation}
where the hadrons $h_1$ and $h_2$ in the final state are produced through the fusion subprocess of two photons emitted by the nuclei: $\gamma_1\,(q_1)\;\gamma_2\,(q_2)\,
\to q(k_q)\;\bar{q}(k_{\bar q}) \to h_1(P_1) \; h_2(P_2) + X$, with $q_1^2\approx q_2^2 \approx 0$. Therefore, the corresponding cross section can be easily obtained by replacing: the flux factor $(1/2s)$ with $(1/2s_{\rm NN})$ in Eq.~(\ref{eq:A3:gensig}), where $s_{\rm NN}$ is the nucleon-nucleon cm energy squared;  the helicity density matrix in Eq.~\eqref{eq:hel-density-matrix} with $\rho_{\lambda,\lambda^\prime}(\gamma_i) = \tfrac12\,\delta_{\lambda,\lambda^\prime}$; $f_{\gamma/\ell}(\xi)$ with $f_{\gamma/{\cal N}}(\xi)$, and $\Delta_L f_{\gamma/\ell}(\xi)$ with 0 in Eqs.~\eqref{eq:dsig-full} and~\eqref{eq:dsig-pt-int}. As a consequence, only the unpolarized WW distribution contributes, and the azimuthal moment $\langle A_{LL}|\cos\phi_{12}\rangle$, Eq.~\eqref{eq:azimom}, induced by longitudinally polarized photons, is absent in UPCs.

\subsection{Weizs\"acker-Williams distributions}

The cross section and the azimuthal moments derived above depend on the colliding beams only through the photon WW distributions, which give the probability density of finding a quasi-real photon carrying a fraction $\xi$ of the parent-particle momentum. Let us then introduce the parametrizations adopted in our analysis, first for unpolarized and longitudinally polarized photons emitted by leptons, and then for unpolarized photons emitted by heavy nuclei in UPCs. In the former case, we will adopt the results of Refs.~\cite{Frixione:1993yw,deFlorian:1999ge}:
\begin{align}
f_{\gamma/\ell}(\xi)\,
&=\,\frac{\alpha}{2\pi}
\left[
\frac{1+(1-\xi)^{2}}{\xi}\,
\ln\!\left(\frac{Q_{\max}^{2}}{Q_{\min}^{2}}\right)
+2m_\ell^{2}\,\xi\!\left(\frac{1}{Q_{\max}^{2}}-\frac{1}{Q_{\min}^{2}}\right)
\right],\label{eq:WWf-frix}\\
\nonumber\\
    \Delta_L f_{\gamma/\ell}(\xi)
&=\,\frac{\alpha}{2\pi}
\left[
\frac{1-(1-\xi)^{2}}{\xi}\,
\ln\left(\frac{Q_{\max}^{2}}{Q_{\min}^{2}}\right)
+2m_\ell^{2}\,\xi^{2}\!\left(\frac{1}{Q_{\max}^{2}}-\frac{1}{Q_{\min}^{2}}\right)
\right]\,,\label{eq:WWdf-frix}
\end{align}
where
\begin{equation}
Q_{\min}^{2}\,=\,\frac{m_\ell^{2}\,\xi^{2}}{1-\xi},\qquad\qquad
Q_{\max}^{2}\,=\,\frac{s}{4}\,\theta_c^{2}\,(1-\xi)+Q_{\min}^{2}\,.
\label{eq:q2mM}
\end{equation}
Here $m_\ell$ is the mass of the beam leptons, $\theta_c$ is the maximum deflection angle of the outgoing lepton from the flight direction of the beam leptons in the cm frame, that depends on the experimental setup. 

\begin{figure}[t]
\centering
\includegraphics[width=8.5cm, keepaspectratio]{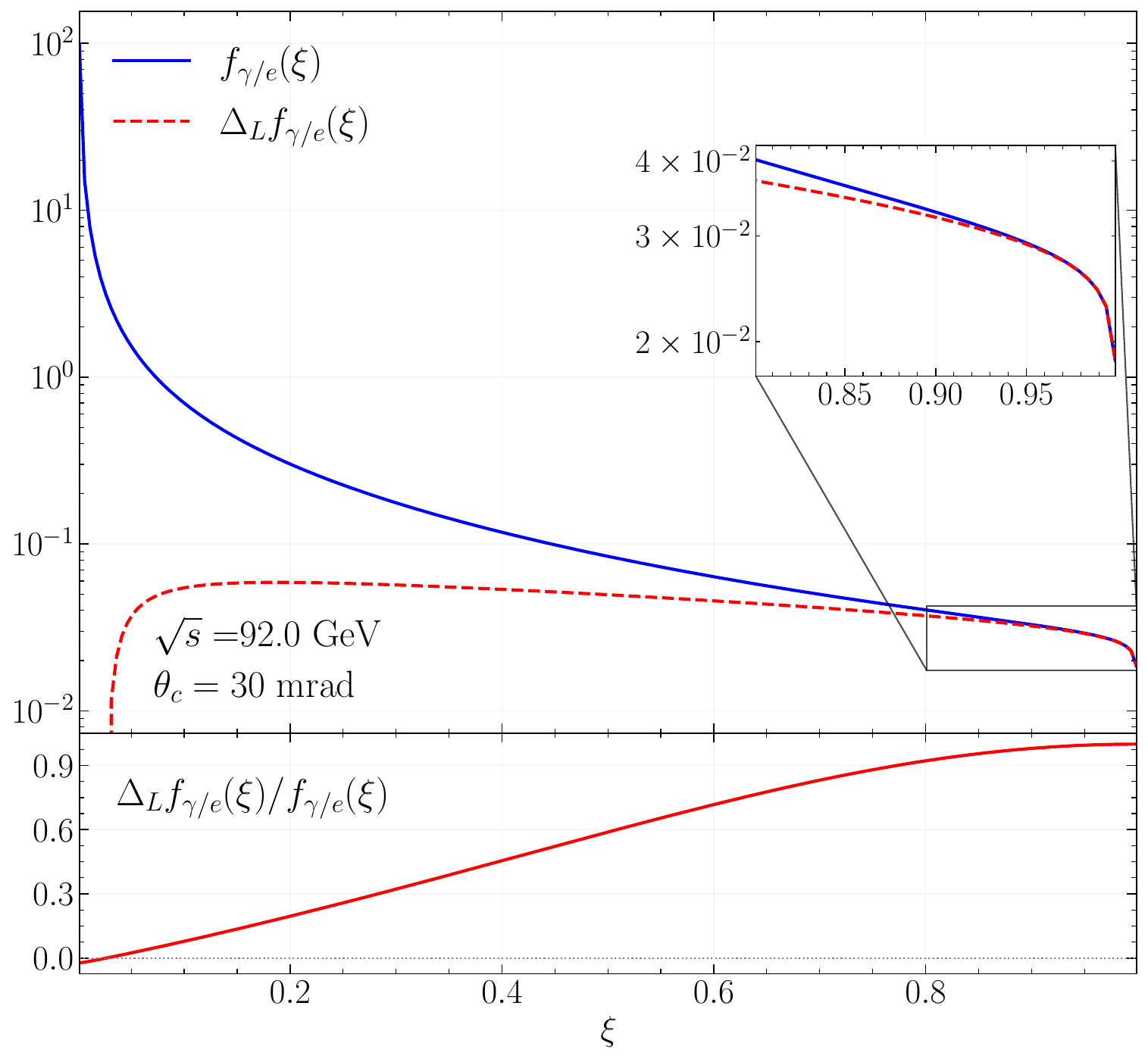}
\caption{Top panel: WW distributions for unpolarized (blue solid line) and longitudinally polarized (red dashed line) photons inside electrons at the FCC-ee energy, $\sqrt{s} = 92$ GeV. Bottom panel: ratio of the longitudinally polarized to the unpolarized photon WW distributions. The value $\theta_c = 30$ mrad is taken from Table 31.1 of Ref.~\cite{ParticleDataGroup:2024cfk}.}
\label{fig:ww-dist-electron}
\end{figure}

In what follows, since we will consider FCC-ee kinematics, we will adopt $\theta_c = 30$ mrad from Table 31.1 of Ref.~\cite{ParticleDataGroup:2024cfk}. Given its relevance for the phenomenological analysis, a comparison between the WW distributions for unpolarized and longitudinally polarized photons inside an electron, Eqs.~(\ref{eq:WWf-frix}) and (\ref{eq:WWdf-frix}), for FCC-ee kinematics is shown in Fig.~\ref{fig:ww-dist-electron}. 

From Fig.~\ref{fig:ww-dist-electron} it is clear that the prefactor in the expression of $\langle\,A_{LL}|\cos\phi_{12}\,\rangle$, Eq.~\eqref{eq:azimom}, given by the ratio of the longitudinally polarized and the unpolarized WW distributions, may be sizable (order of unity) only at very large values of $\bar \xi_{1,2}$, where the corresponding cross sections will be strongly suppressed.

Since in the sequel we will also consider the case of heavy-nuclei ultra-peripheral collisions, we need an explicit expression also for the unpolarized photon WW distribution inside nuclei. We will adopt the following form (see, e.g.,~Ref.~\cite{Jackson:1998nia})
\begin{align}
f_{\gamma/{\cal N}}(\xi) = \frac{Z^2 \alpha}{\pi} \, \frac{1}{\xi}\, \bigg [ 2 x_i\, K_0(x_i) K_1(x_i) \,-\, x_i^2 \bigl(\,K_1^2(x_i) - K_0^2(x_i)\,\bigr) \bigg ] \, , 
\label{eq:EPA_nuclei}
\end{align}
where $\xi$ is the fraction of the momentum of a charged nucleon inside the nucleus carried by the photon. The above expression involves an integration over the impact parameter $b$ between a given $b_\text{min}$ and infinity, where $b$ is the transverse  distance (w.r.t. the beam axis) between the two colliding nuclei. In Eq.~\eqref{eq:EPA_nuclei}, $x_i = \xi \, M_N \, b_\text{min}$, with $M_N$ being the nucleon mass, whereas  $K_0$, $K_1$ are the zeroth- and first-order modified Bessel functions of the second kind, related respectively to the emission of longitudinally and transversely polarized photons. The transverse polarization component dominates for ultrarelativistic particles, which have a Lorentz factor $\gamma \gg 1$.

The integral over the impact parameter $b$ and the value of $x_i$ depend on the choice of $b_{\rm min}$. For symmetric heavy-ion collisions, $b_{\rm min} = 2 R_{\cal N}$, with $R_{\cal N}$ being the radius of the nucleus $\cal N$. In what follows, we adopt the following simple functional form:
\begin{equation}
        R_{\cal N} = R_0 A^{1/3}\,,
\end{equation}
where $R_0 \simeq 1.2$ fm is the Fermi radius and $A$ is the atomic number of the nucleus $\cal N$.

%% file: Phenomenology.tex
\section{Phenomenology}\label{sec:pheno}

In this section we present phenomenological results for almost back-to-back pion-pair production in quasi-real photon-photon collisions. We first summarize the adopted parametrizations for the unpolarized and Collins TMD FFs and then present our results for different future experiments at $e^+e^-$ and $\cal NN$ colliders.

\subsection{TMD parametrizations}\label{sec:parametrizations}

In what follows, we will adopt different parametrizations of the collinear unpolarized pion fragmentation functions and of the first $p_\perp$-moment of the Collins TMD FFs. To this end, we will employ two recent phenomenological extractions available in the literature~\cite{Boglione:2024dal,Gamberg:2022kdb}. They are similar in approach but differ in some details, as illustrated below. Both of them are based on global fits of data on the Collins azimuthal asymmetry measured in SIDIS and $e^+e^-$ SIA processes, and employ a factorized $z$- and $p_\perp$-dependence for the functional form of the TMD FFs. The $p_\perp$ dependence of the unpolarized TMD FF is parametrized through Gaussian factors of the type
\begin{eqnarray}
{\cal G}(p_\perp^2)
&=& \frac{1}{\pi\avpperp}\;
{\exp\!\left[{-\frac{p_\perp^2}{\avpperp}}\right]},
\label{eq:Ggauss}
\end{eqnarray}
while the one of the Collins FF is proportional to the same factor, multiplied by an additional linear $p_\perp$ term.
The full Collins-Soper-Sterman formalism for the TMD evolution with the energy scale is not implemented at this level. However, the collinear DGLAP evolution is accounted for in the factors proportional to the collinear FF.

Although the full TMD scale evolution should be taken into account, there is at present a general consensus in the literature, based on experimental evidence, that these effects appear to be mild for observables such as spin and azimuthal asymmetries. In fact, since these quantities are defined as ratios of cross sections, evolution and higher order effects tend to partially cancel out~\cite{Kang:2017btw}. It has been shown that in different processes this kind of parametrization is compatible with full TMD evolution at higher logarithmic accuracy~\cite{Kang:2017btw,DAlesio:2017bvu} (see e.g.~Fig.~14 of Ref.~\cite{STAR:2017akg}). This aspect requires further studies and TMD evolution should be taken into proper account in future phenomenological analyses.

The first parametrization adopted, taken from Ref.~\cite{Boglione:2024dal}, will be dubbed "TO-CA24" in what follows. The unpolarized TMD fragmentation function is parametrized as %
\begin{equation}
D_{h/q}(z,p_\perp)\,=\,D_{h/q}(z)\,{\cal G}(p_\perp^2)\,,
\label{eq:unp-FF-par}
\end{equation}
with a single, flavor- and scale-independent Gaussian width $\langle\,p_\perp^2\rangle \,=\,0.12$ GeV$^2$,
as extracted from a fit to HERMES multiplicities~\cite{Anselmino:2013lza}, and adopting the set of collinear pion FFs, $D_{h/q}(z)$, by de Florian, Sassot and Stratmann (DSS)~\cite{deFlorian:2007aj}.

As said, the Collins function is taken proportional to the corresponding unpolarized TMD FF times a factorized, Gaussian-like $p_\perp$-dependent factor:
\begin{equation}
\Delta^N D_{h/q^\uparrow}(z,p_\perp)\,\equiv \frac{2p_\perp}{z m_h}H_{1\,h/q}^\perp(z,p_\perp)\,=\,{\cal N}_q^C(z)\,\frac{p_\perp}{M_C}\,2\,\sqrt{2\,e}\,e^{- p_\perp^2/M_C^2}\,D_{h/q}(z,p_\perp)\,,
\label{eq:col-FF-par}
\end{equation}
where $H_{1\,h/q}^\perp(z,p_\perp)$ is the Collins function in the Amsterdam notation,  $q={\rm fav},\,{\rm unf}$ (favored, e.g.~$u\to\pi^+$, or unfavored, e.g.~$d\to\pi^+$, FFs) and $M_C$ is a free parameter with mass dimension.
By requiring that $|{\cal N}_q^C(z)|\leq 1$ this functional form guarantees that the positivity bound for the Collins function is fulfilled separately for the $z$ and $p_\perp$ components of the FF and, for the latter, is saturated at its maximum value.
Specifically, the ${\cal N}_q^C(z)$ factors for the parametrization of Ref.~\cite{Boglione:2024dal} read
\begin{equation}
    {\cal N}_{\rm fav}^C(z)\,=\,N_{\rm fav}^C\,z^\gamma,\qquad {\cal N}_{\rm unf}^C(z)\,=\,N_{\rm unf}^C\,.
\end{equation}

The values of the four parameters of the TO-CA24 fit adopted here are: %$\langle p_\perp^2\rangle$, 
$N^C_{\rm fav}= 0.86 \pm 0.06$, $N^C_{\rm unf}= -0.35 \pm 0.02$, $\gamma= 1.75 \pm 0.16$, and $M_C^2 = (0.26 \pm 0.03)$ GeV$^2$, with uncertainties at $1\sigma$ confidence level.

The second extraction considered is the JAM3D-2022 set, Ref.~\cite{Gamberg:2022kdb}. Here, on top of SIDIS and $e^+e^-$ SIA data, the fit procedure includes also data on the transverse single-spin asymmetry $A_N(p^\uparrow p \to \pi + X)$ for inclusive pion production in $pp$ collisions, 
with or without the inclusion of lattice QCD results.
The fit without lattice input is adopted here and labeled "\JAM3D" in what follows. We have nevertheless checked that the extraction including lattice results leads to estimates similar to the ones based on the \JAM3D extraction.

Also in the JAM case a Gaussian ansatz is used for the unpolarized TMD FF, with the LO DSS pion FFs~\cite{deFlorian:2007ekg} as collinear input. Moreover, two different unpolarized Gaussian widths are employed for the favored and unfavored cases, $\avpperp_{\rm fav} = 0.125$ GeV$^2$ and $\avpperp_{\rm unf} = 0.146$ GeV$^2$ respectively.

The Collins function is given in terms of its first $p_\perp$ moment as 
\begin{eqnarray}
H_{1\,  h/q}^{\perp}(z,p_\perp^2)
&=& \frac{2 z^2 m_h^2}{\avpperp^{h/q}_{H_1^\perp}}\,
    H_{1\, h/q}^{\perp (1)}(z)\
    {\cal G}_{H_1^\perp}^{h/q}(p_\perp^2)\,,\quad (q = {\rm fav}, {\rm unf})\,,
\label{eq:collins-JAM}
\end{eqnarray}
where the factor ${\cal G}_{H_1^\perp}^{h/q}(p_\perp^2)$ has the same functional form as in Eq.~(\ref{eq:Ggauss}), with $\langle\,p_\perp^2\,\rangle$ replaced by $\langle\,p_\perp^2\,\rangle_{H_1^\perp}^{h/q}$, the Collins Gaussian width. The Collins first $p_\perp$ moment is parametrized as 
\begin{equation}
H^{\perp(1)}_{1\,h/q}(z)\!=\!\frac{N_q\,z^{a_q}(1-z)^{b_q}(1+\gamma_q\,(1-z)^{\beta_q})}
            {{\rm B}[a_q\!+\!2,b_q\!+\!1]
             +\gamma_q {\rm B}[a_q\!
             +\!2,b_q\!
             +\!\beta_q\!+\!1]}   
\,,
\label{eq:col-first-mom-JAM}
\end{equation}
where $N_q$, $a_q$, $b_q$, $\gamma_q$ and $\beta_q$ are free parameters and B is the Euler beta function.

The relation between the Collins first $p_\perp$ moment in the Amsterdam notation and the quantity $\Delta^N D_{h/q^\uparrow}(z)$ entering Eqs.~(\ref{eq:dsig-pt-int}) and (\ref{eq:azimom}) reads
\begin{equation}
H_{1\, h/q}^{\perp (1)}(z) = \frac{ \sqrt{\langle p_\perp^2\rangle_C}}{2\sqrt{\pi} z m_h} \Delta^N D_{h/q^\uparrow}(z)\,,
\label{eq:H1-DeltaN}
\end{equation}
where $\langle p_\perp^2\rangle_C = M_C^2\,\langle p_\perp^2\rangle / (M_C^2 + \langle p_\perp^2\rangle)$.

A comparison of the first $p_\perp$ moments of the Collins function from the TO-CA24 and JAM3D-2022 extractions at $Q^2=4$ GeV$^2$ is presented in \cf{fig:Coll-first-mom-comp}. As one can see, the two JAM3D-2022 extractions are very similar to each other, while the TO-CA24 one differs both in magnitude and in the uncertainty size, especially for the favored case at $z \lesssim 0.4$. As we will see later, this translates into some differences in the predictions based on the two extractions.

\begin{figure}[t]
    \centering
    \includegraphics[width=9cm, keepaspectratio]{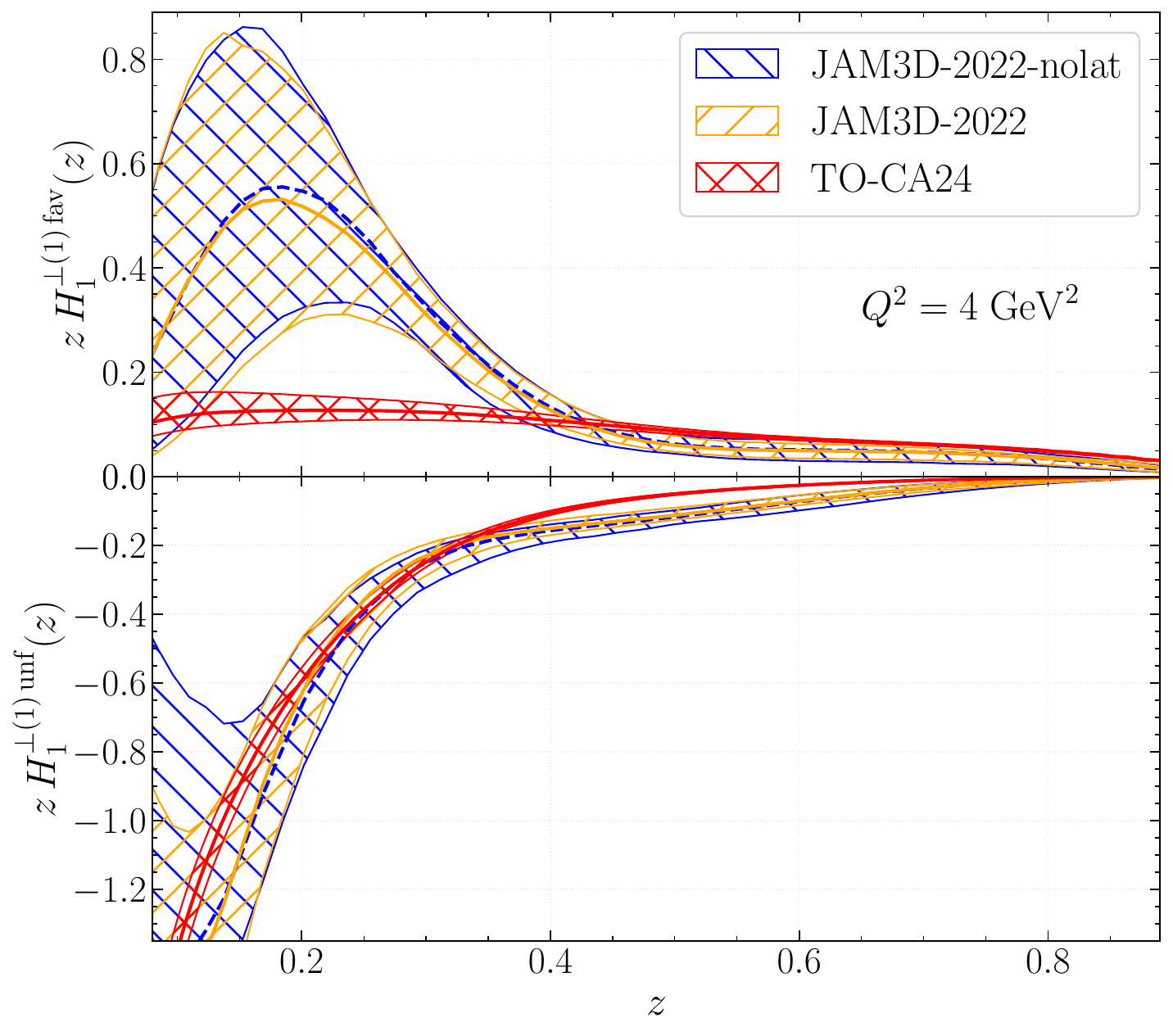}
    \caption{Comparison of the Collins first $p_\perp$ moments for the favored (upper panel) and unfavored (lower panel) case from the TO-CA24~\cite{Boglione:2024dal} (red bands and curves) and JAM3D-2022~\cite{Gamberg:2022kdb} (orange and blue (nolat) bands and curves) analyses at $Q^2 = 4$ GeV$^2$. Uncertainty bands represent the corresponding 1$\sigma$ confidence level.}
    \label{fig:Coll-first-mom-comp}
\end{figure}

\subsection{Results}\label{sec:results}

We now present our predictions for the unpolarized cross sections and Collins azimuthal asymmetries in $e^+e^- ({\cal NN}) \to  h_1 h_2\, X$ through the subprocess 
$\gamma_1 \gamma_2 \to q \bar q$. As already mentioned, we will concentrate on light-hadron production, in particular charged pions. As is customary in $e^+e^- \to h_1 h_2 X$ SIA processes, we introduce unlike (different charge) and like (same charge) sign combinations of pion pair production. For instance, in the case of unpolarized cross sections and the denominator of the asymmetries, we consider the combinations:
\begin{equation}
    U = D_{\pi^+/q}(z_1) D_{\pi^- /\bar{q}} (z_2) + D_{\pi^-/q}(z_1) D_{\pi^+ /\bar{q}} (z_2)\,,\qquad  %\nonumber\\
    L = D_{\pi^+/q}(z_1) D_{\pi^+ /\bar{q}} (z_2) + D_{\pi^-/q}(z_1) D_{\pi^- /\bar{q}} (z_2)\,. 
\label{eq:U-L-comb}
\end{equation}
The same combinations for the numerator of the azimuthal asymmetries are obtained by the exchange $D_{\pi/q}(z) \to \Delta^N D_{\pi/q^\uparrow}(z)$.

We will consider $e^+ e^-$ collisions in the kinematic configuration of the planned FCC-ee~\cite{ParticleDataGroup:2024cfk,FCC:2025lpp} and PbPb collisions for the future High-Luminosity phase at the LHC (HL-LHC)~\cite{ATLAS:2023dns,ATLAS:2024mvt,CMS:2023gfb,ATLAS:2025aav,CMS:2022lbi,CMS:2025lsm,ALICE:2023udb,LHCb:2023hlw,LHCb:2022ahs} at the different LHC detectors. Apart from the initial colliding beams, these experiments differ in the $\sqrt{s}$ value and, for the LHC case, in the covered pseudorapidity range, determined by the specific features of the detectors. Hence, we will adopt different specific kinematical cuts for our predictions, as summarized in Table~\ref{tab:expt-cuts}.

\begin{table}[ht]
    \centering
    \begin{tabular}{ccccc}
    \toprule
        Experiment & Collision type & 
        $\sqrt s\;\;[{\rm GeV}]$ 
        & $\int {\cal L}dt\;\;[{\rm fb}^{-1}/{\rm month}]$ &  Kinematical cuts \\
        \midrule
        \makecell[c]{FCC-ee\\ \cite{ParticleDataGroup:2024cfk,FCC:2025lpp}} & $e^+e^-$ & 92 & $1.4\cdot10^3$ & \makecell[c]{$\lvert\eta_q,\eta_{\bar q}\rvert \leq 2.5 (0.5)$ \\[2mm] 
        $k_T \in [3,20]\, ([20,40])$ GeV\\[2mm]} \\
        \midrule        
        \makecell[c]{ATLAS/CMS \\ \cite{ATLAS:2023dns,ATLAS:2024mvt,ATLAS:2025aav,CMS:2023gfb,CMS:2025lsm,CMS:2022lbi}} & PbPb & 5360 & $2.1\cdot 10^{-6}$ & \makecell[c]{$\lvert\eta_q,\eta_{\bar q}\rvert \leq 2.4$ \\[2mm] $k_T \in [10,30]$ GeV\\[2mm]} \\
        \midrule
        \makecell[c]{ALICE\\ \cite{ALICE:2023udb}} & PbPb & 5360 & $2.5\cdot 10^{-6}$  &\makecell[c]{$\lvert\eta_q,\eta_{\bar q}\rvert \leq 0.9$ \\[2mm] $k_T \in [10,30]$ GeV\\[2mm]} \\
        \midrule
        \makecell[c]{LHCb\\ \cite{LHCb:2022ahs,LHCb:2023hlw}} &  PbPb & 5360 & $0.5\cdot 10^{-6}$  & \makecell[c]{$2 < \eta_q,\eta_{\bar q} < 4.5$ \\[2mm] $k_T \in [10,30]$ GeV\\[2mm]} \\ 
    \bottomrule
    \end{tabular}
    \caption{Summary of the kinematical configurations and cuts considered for our predictions at $e^+e^-$ and heavy-ion colliders. The values of the integrated luminosity are taken from Ref.~\cite{ParticleDataGroup:2024cfk} for FCC-ee and Refs.~\cite{dEnterria:2022sut,Citron:2018lsq} for LHC.}
    \label{tab:expt-cuts}
\end{table}

\vspace*{-1mm}
\subsubsection{$e^+e^-$ colliders}

\begin{figure}[H]
\centering
\includegraphics[width=8.5cm, keepaspectratio]{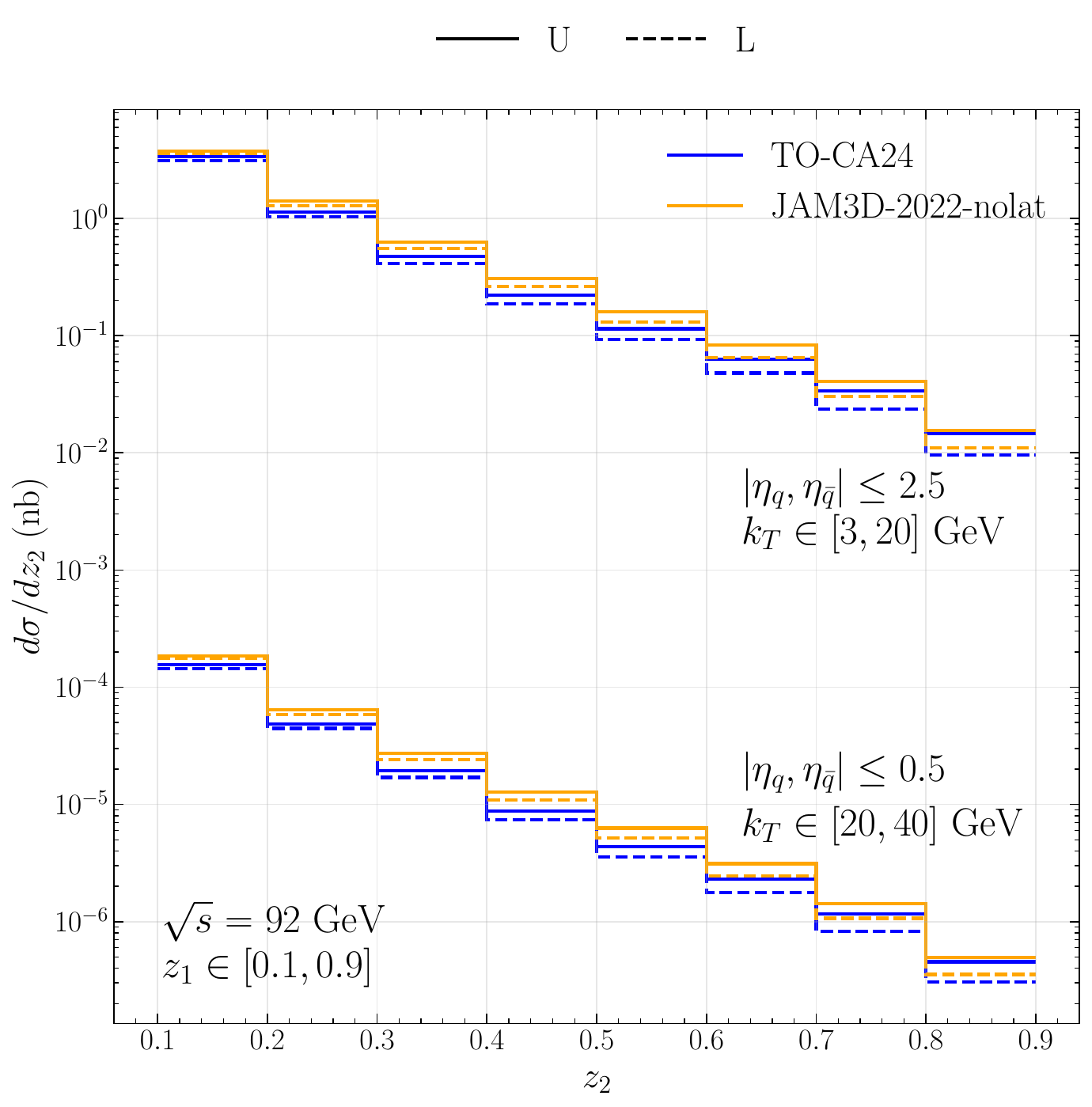}
\caption{Unpolarized differential cross section for the FCC-ee kinematical setups of Table~\ref{tab:expt-cuts} at $\sqrt{s} = 92$ GeV, as a function of $z_2$ and integrated over bins of width $\Delta z_2 = 0.1$. Predictions are based on the \TOCA (blue lines) and \JAM3D (orange lines) extractions for the unlike-sign (U, solid lines) and like-sign (L, dashed lines) combinations of Eq.~\eqref{eq:U-L-comb}.}
\label{fig:FCC-ee-dsigma}
\end{figure}

Let us start with the predictions for FCC-ee. In Fig.~\ref{fig:FCC-ee-dsigma} we present the unpolarized differential cross section of Eq.~\eqref{eq:dsig-pt-int} as a function of $z_2$, integrated over bins of width $\Delta z_2 = 0.1$, and integrated over $z_1$ in the range $[0.1:0.9]$ and over $k_T$ and $\eta_{q,\bar q}$ in two different selections of kinematical cuts, as shown in the plot legends and listed in Table~\ref{tab:expt-cuts}.  The predictions are computed using the unpolarized TMD FFs of the \TOCA (blue lines) and \JAM3D (orange lines) extractions, and are presented for the unlike-sign (U, solid lines) and like-sign (L, dashed lines) combinations given in Eq.~\eqref{eq:U-L-comb}. The kinematical configuration with a narrower $\eta_{q,\bar q}$ and a larger $k_T$ range (lower set of curves in Fig.~\ref{fig:FCC-ee-dsigma}) is selected to explore a kinematical domain where the azimuthal moment $\langle A_{LL}\vert \cos\phi_{12}\rangle$ may be larger. In fact, within this range one probes larger $\xi$ values in the WW distributions ($\xi \gtrsim 0.7$), where the longitudinally polarized photon distribution becomes comparable with the unpolarized one, as shown in Fig.~\ref{fig:ww-dist-electron} (small inset and lower panel).

As one can see from Fig.~\ref{fig:FCC-ee-dsigma}, the cross section decreases at larger values of $z_2$. In the case of the full $\eta_{q,\bar q}$ range, $|\eta_{q,\bar q}|\leq 2.5$ (upper set of curves), it spans from ${\cal O}(10^0)$ nb at small $z_2$ to ${\cal O}(10^{-2})$ nb at large $z_2$, while in the narrower $\eta_{q,\bar q}$ interval ($|\eta_{q,\bar q}|\leq 0.5$, lower set of curves), it goes from ${\cal O}(10^{-4})$ to ${\cal O}(10^{-6})$ nb within the same $z_2$ range. Overall, the cross section is similar for the unlike-sign and like-sign combinations at low $z_2$ values, while it is larger at large $z_2$ for the unlike-sign case. Moreover, the predictions are slightly larger for the \JAM3D case (orange lines) in both kinematical ranges considered.

\begin{figure}[htbp]
\centering
\subfloat[]{\label{fig:FCC-ee-kT-3-20}
\includegraphics[width=8.9cm, keepaspectratio]{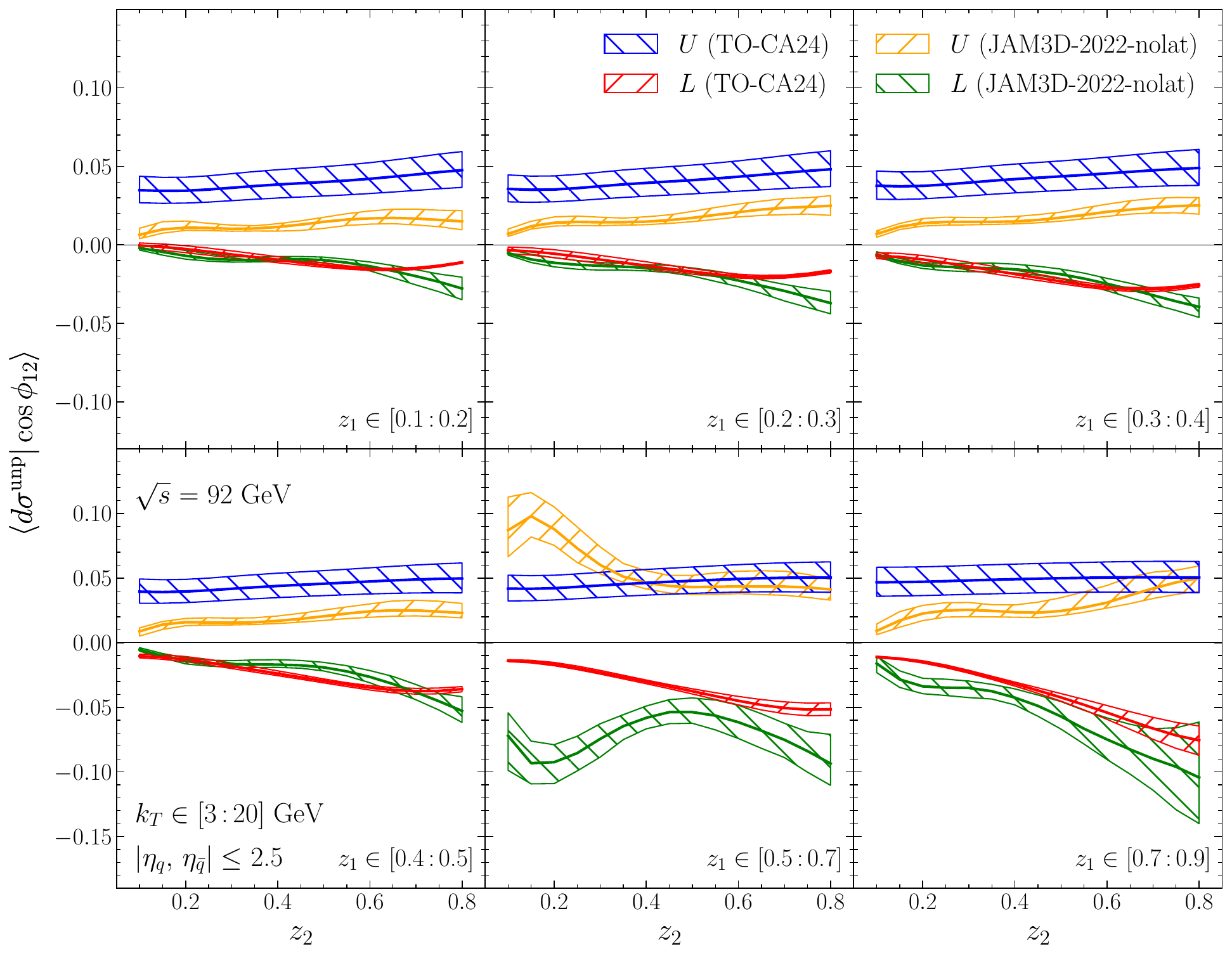}
}
\subfloat[]{\label{fig:FCC-ee-kT-20-40}\includegraphics[width=8.9cm, keepaspectratio]{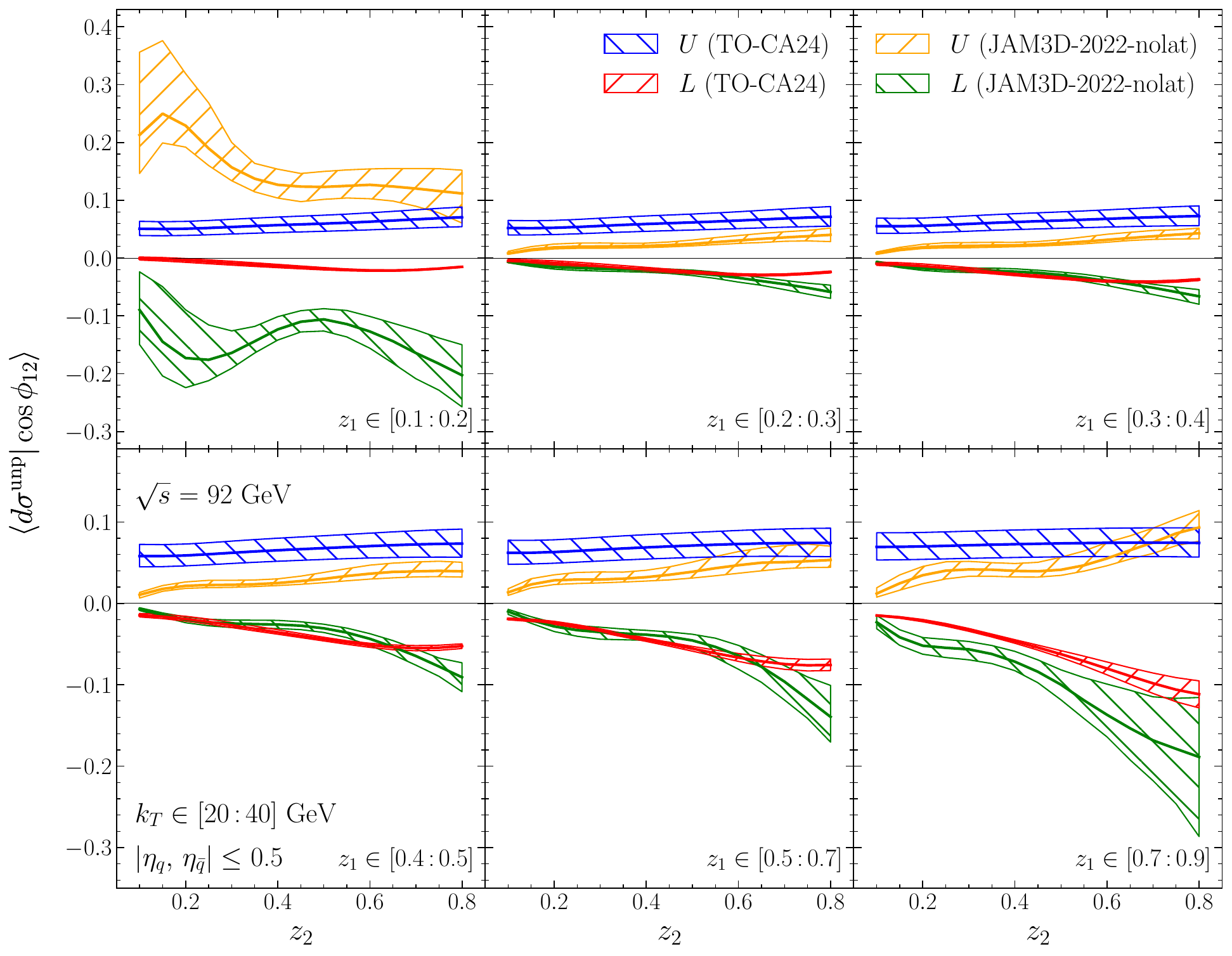}
}
\caption{The azimuthal moment $\langle {\rm d}\sigma^{\rm unp} \vert \cos\phi_{12} \rangle$  for FCC-ee  at $\sqrt{s} = 92$ GeV, as a function of $z_2$ in six bins of $z_1$ and integrated over: (a) $|\eta_{q,\bar q}|\leq 2.5$, $3 \leq k_T \leq 20$ GeV; (b) $|\eta_{q,\bar q}|\leq 0.5$, $20 \leq k_T \leq 40$ GeV. Predictions are based on the \TOCA (blue and red bands) and \JAM3D (orange and green bands) extractions for the unlike-sign (blue and orange bands) and like-sign (red and green bands) combinations of Eq.~\eqref{eq:U-L-comb}. Uncertainty bands represent the corresponding 1$\sigma$ confidence level. The central solid lines with the same color codes represent the median of the predictions based on the two extractions. Notice the different scale in the vertical axes for the left and right plots.}
\label{fig:FCC-ee-asym}
\end{figure}

\begin{figure}
\includegraphics[width=8.9cm, keepaspectratio]{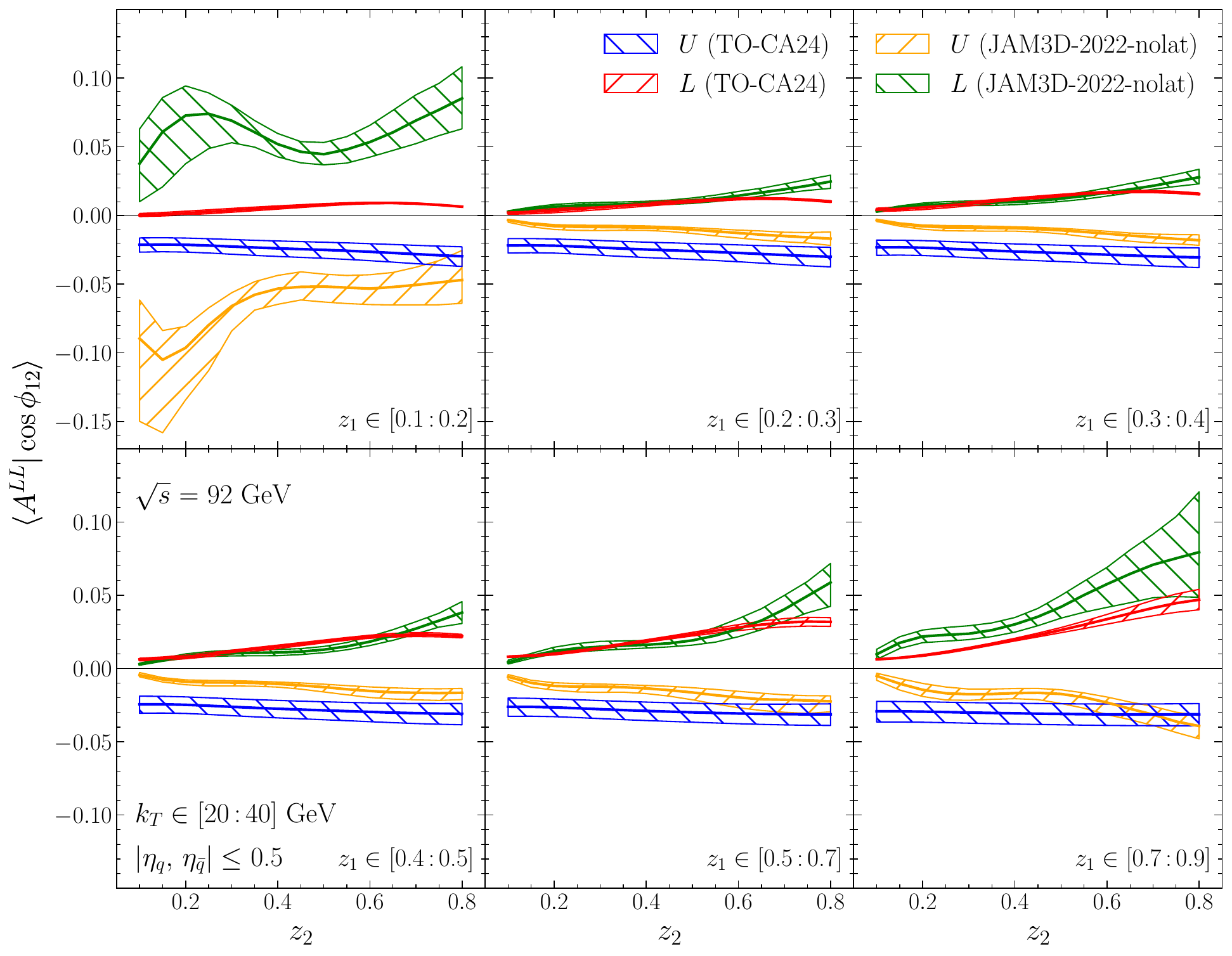}
\caption{$\langle A_{LL} \vert \cos\phi_{12} \rangle$ azimuthal moments for FCC-ee at $\sqrt{s} = 92$ GeV as a function of $z_2$ in six bins of $z_1$ and integrated over $|\eta_{q,\bar q}|\leq 0.5$ and $20 \leq k_T \leq 40$ GeV. Predictions are based on the \TOCA (blue and red bands) and \JAM3D (orange and green bands) for the unlike-sign (blue and orange bands) and like-sign (red and green bands) combinations of Eq.~\eqref{eq:U-L-comb}. Uncertainty bands represent the corresponding 1$\sigma$ confidence level. The central solid lines with the same color codes represent the median of the predictions based on the two extractions. }
\label{fig:FCC-ee-ALL-asym}
\end{figure}

In Fig.~\ref{fig:FCC-ee-asym} we present our predictions for the azimuthal moment of the unpolarized cross section,  $\langle {\rm d}\sigma^{\rm unp} \vert \cos\phi_{12}\rangle$, as a function of $z_2$ and in six bins of $z_1$, for the FCC-ee setup at $\sqrt{s}=92$ GeV and the two different sets of kinematical ranges for the transverse momentum $k_T$ and the rapidities considered in Table~\ref{tab:expt-cuts} (Figs.~\ref{fig:FCC-ee-kT-3-20} and \ref{fig:FCC-ee-kT-20-40} respectively). In analogy with the unpolarized cross section case of Fig.~\ref{fig:FCC-ee-dsigma}, we present the azimuthal moments for the unlike-sign (blue and orange bands) and like-sign (red and green bands) combinations, adopting both the \TOCA (blue and red bands) and the \JAM3D (orange and green bands) extractions. The median of the predictions based on 
the two extractions are also shown as solid lines and with the same color codes.

Let us comment on these results. Our estimates for the $\langle {\rm d}\sigma^{\rm unp} \vert \cos\phi_{12}\rangle$ azimuthal moment turn out to be positive for the unlike-sign combination and negative for the like-sign one for both the extractions considered. Within uncertainties, for the \JAM3D extraction, apart from the mentioned different sign, they are on average comparable in size but slightly larger for the like-sign combination in the large $z_1$ region. More specifically, for the unlike-sign combination (orange bands) the azimuthal moment is on average of the order of 2-3\%, increasing up to about 5-10\% at large values of $z_1$ and $z_2$. As for the like-sign combination (green bands), it is of order 5-10\% in size, shows a similar specular behavior as a function of $z_2$ and tends to slightly increase in size at the largest $z_1$ bins, up to 10-20\%. Concerning the \TOCA extraction, it shows in general a smoother behavior. In this case, the unlike-sign combination (blue bands) is positive and on average larger in size than the negative, like-sign one (red bands), although at the largest $z_{1,2}$ values their size becomes more similar, ranging from a few percent up to about 10\% at most in the largest $z_2$ region. Moreover, for the like-sign combination the central values of the \JAM3D and \TOCA extractions (green and red lines respectively) are in general quite similar in size, except for few bins where the \JAM3D extraction leads to a larger size, even if with bigger uncertainties. Concerning the unlike-sign combination, the central values of the \JAM3D extraction (orange lines) are in general smaller in size as compared to those of the \TOCA case (blue lines), apart again from few bins.

We also note that the uncertainty bands for the \JAM3D extraction are much larger than those for the \TOCA ones. As already discussed, this reflects the fact that the \JAM3D extraction includes also $A_N$ data for inclusive pion production in polarized proton-proton collisions, where additional mechanisms other than the Collins effect may play a role. Furthermore, the \JAM3D analysis involves more complicated parametrizations with a larger number of free parameters. Finally, the azimuthal moments of Fig.~\ref{fig:FCC-ee-kT-20-40}, at more central rapidities and much larger $k_T$ values, are on average larger in size as compared to those in the configuration of Fig.~\ref{fig:FCC-ee-kT-3-20} (notice the different scale in the vertical axes), although the corresponding cross sections are expected to be much smaller, as shown in Fig.~\ref{fig:FCC-ee-dsigma}.

At FCC-ee it would also be interesting, although more challenging, to probe azimuthal asymmetries induced by longitudinally polarized photons. Therefore, in Fig.~\ref{fig:FCC-ee-ALL-asym} we present results for the $\langle A_{LL} \vert \cos\phi_{12}\rangle$ azimuthal moment, Eq.~\eqref{eq:azimom}, in the large $k_T$ range and at central rapidity. In this case, we do not show the corresponding results within the full rapidity range at smaller $k_T$ values, as the asymmetry turns out to be very small, of the order of a few per mille. Apart from the overall sign (see Eq.~\eqref{eq:azimom}) , the $\langle A_{LL} \vert \cos\phi_{12}\rangle$ moment exhibits qualitative features similar to those of the $\langle {\rm d}\sigma^{\rm unp} \vert \cos\phi_{12}\rangle$ one in the same kinematical setup, Fig.~\ref{fig:FCC-ee-kT-20-40}, although they are in general smaller in size, by approximately a factor of two. 

Therefore, most of the remarks regarding Fig.~\ref{fig:FCC-ee-kT-20-40} also apply, with minor adjustments, to Fig.~\ref{fig:FCC-ee-ALL-asym}. The \TOCA extraction predicts these asymmetries to be of the order of few percent in size, while the \JAM3D based estimates can be as large as 10\% in size, though with larger uncertainties.

These predictions show the potential of FCC-ee to provide relevant complementary information on the Collins fragmentation function and its parametrizations in $\gaga$ collisions. It could provide new precise data at different energy scales and with different flavor combinations as compared to SIDIS and $e^+e^-$ SIA processes, contributing to improving the current phenomenological extractions and discriminating between models and approaches to the TMD quark fragmentation functions.

\subsubsection{UPCs at the LHC}

\begin{figure}[t]
\centering
\includegraphics[width=8.5cm, keepaspectratio]{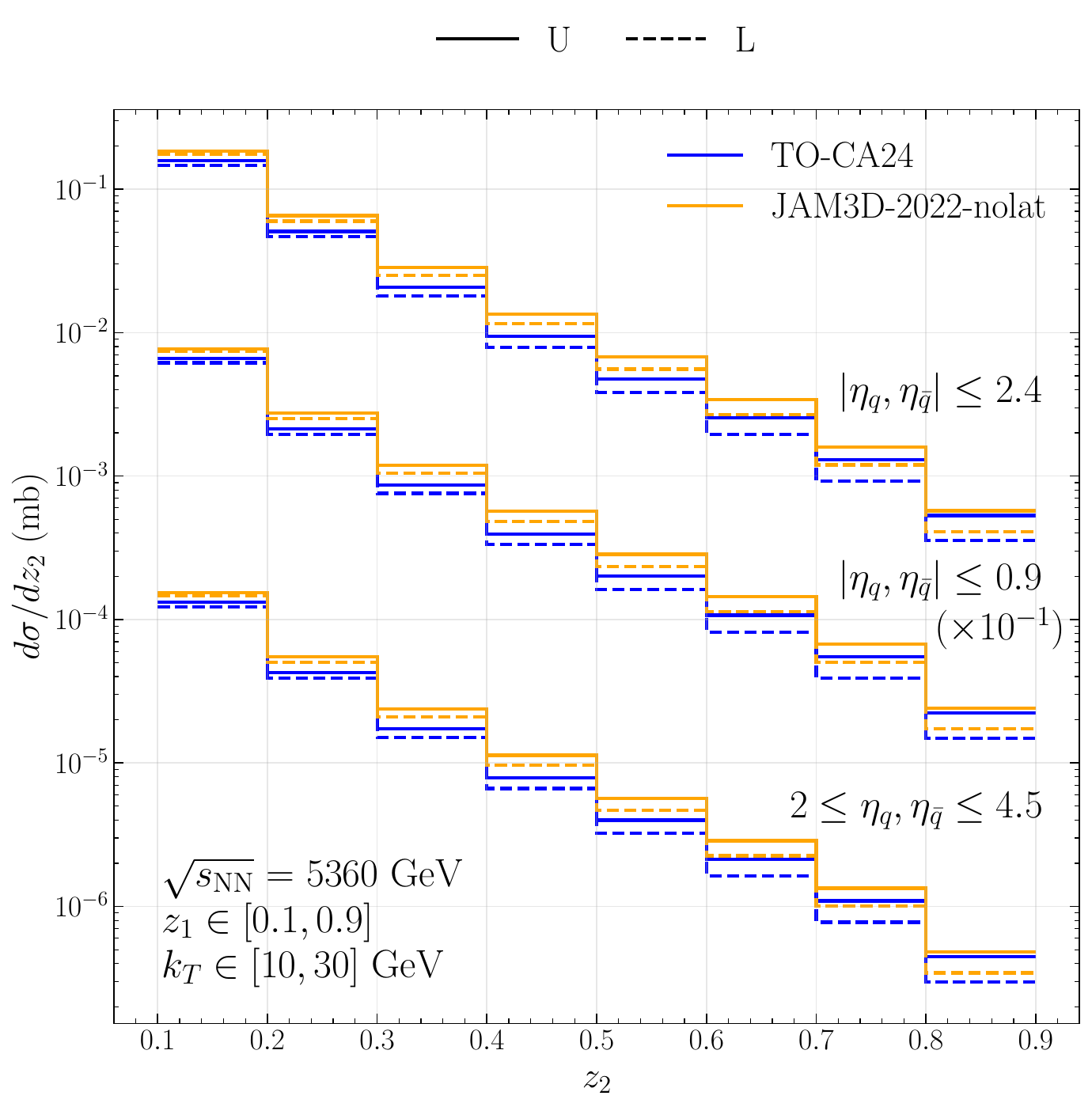}
\caption{Unpolarized differential cross section for PbPb UPCs at the LHC as a function of $z_2$ and integrated over bins of width $\Delta z_2=0.1$, for the three different kinematical configurations (ATLAS/CMS, ALICE and LHCb) summarized in Table~\ref{tab:expt-cuts}. The ALICE cross section is scaled by a factor 0.1 for clarity. The color code for the curves is the same adopted in Fig.~\ref{fig:FCC-ee-dsigma}.}
\label{fig:UPC-HL-LHC-dsigma}
\end{figure}

We now move to the case of UPCs at the LHC. We concentrate here on PbPb collisions, which have the advantage of a large value of the charge number, $Z = 82$, leading to a large enhancement factor $Z^4$ %= 6724$
in the cross section. We consider three different kinematical configurations: one for the ATLAS and CMS detectors, whose pseudorapidity coverage is similar to the one considered for FCC-ee ($\lvert \eta_{q}, \eta_{\bar q}\rvert < 2.4$), one for the ALICE detector, at central rapidity ($\lvert \eta_{q}, \eta_{\bar q}\rvert < 0.9$), and finally one for the LHCb apparatus, at forward rapidity ($2 <\eta_{q}, \eta_{\bar q} < 4.5$). For all of them, we consider the jet transverse momentum in the region $k_T \in [10:30]$ GeV.
The predictions for the $z_2$-differential cross section  within these kinematical domains, integrated over bins of width $\Delta z_2 = 0.1$ and over $z_1 \in [0.1:0.9]$, are presented in Fig.~\ref{fig:UPC-HL-LHC-dsigma} at $\sqrt{s_{\rm NN}} =5.36$ TeV. Again, they are shown for both the \TOCA and \JAM3D extractions and for the unlike-charge and like-charge combinations. Notice also that the predictions for the ALICE configuration are scaled down by a factor of $10^{-1}$ for visual purposes. As expected, the cross section presents the same features as the one for the FCC-ee case in Fig.~\ref{fig:FCC-ee-dsigma}. For the ATLAS/CMS and ALICE configurations it has a similar magnitude,  ranging from about 0.1 mb at low $z_2$ to ${\cal O}(10^{-3}-10^{-4})$ mb at large $z_2$,  while at LHCb, being at forward rapidity, the cross section decreases, from ${\cal O}(10^{-4})$ to ${\cal O}(10^{-6}-10^{-7})$ mb within the considered $z_2$ range.

\begin{figure}[ht]
\centering
\subfloat[ATLAS/CMS]{\label{fig:UPC-CMS}\includegraphics[width=8.9cm, keepaspectratio]{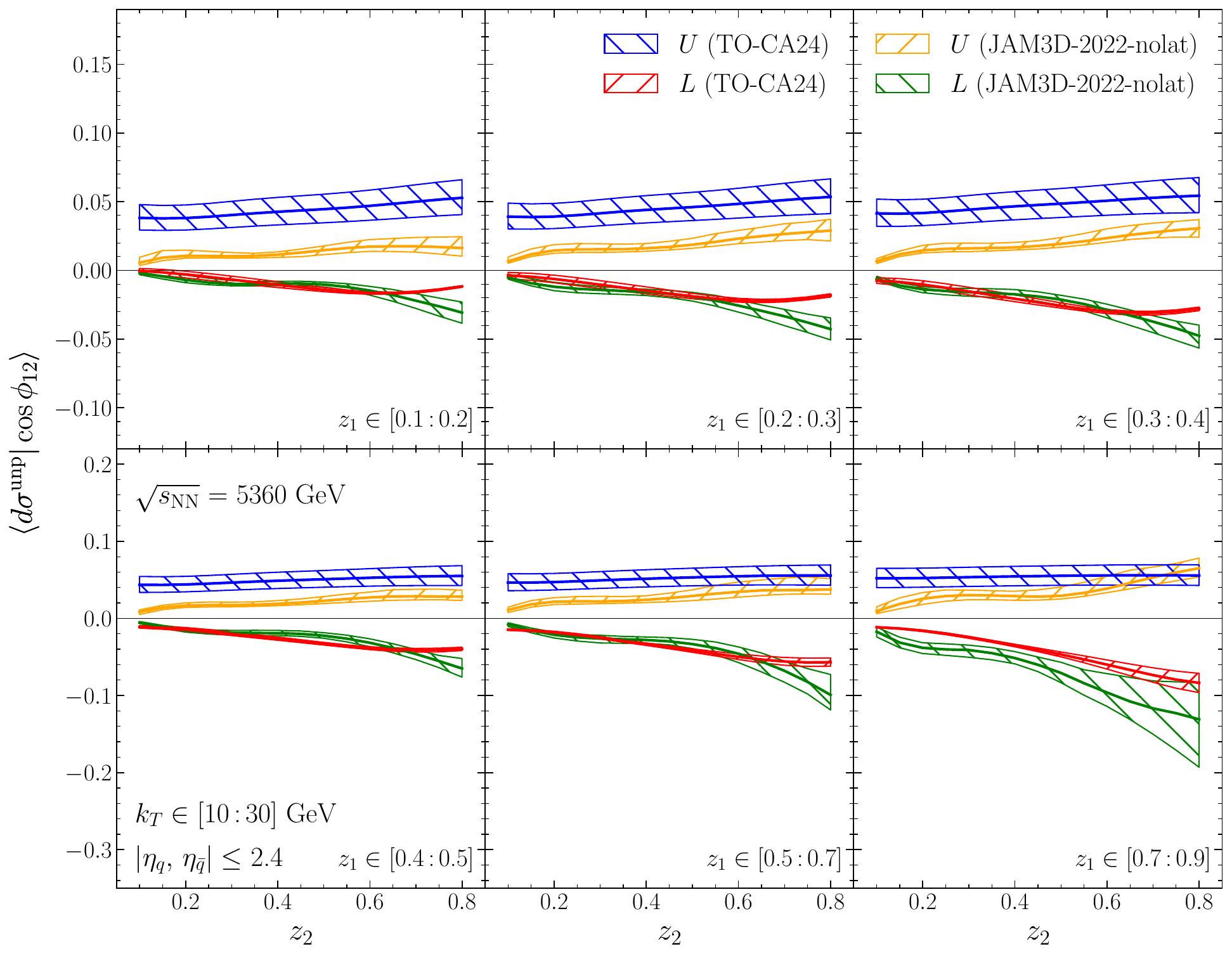}
}
\subfloat[ALICE]{\label{fig:UPC-ALICE}\includegraphics[width=8.9cm, keepaspectratio]{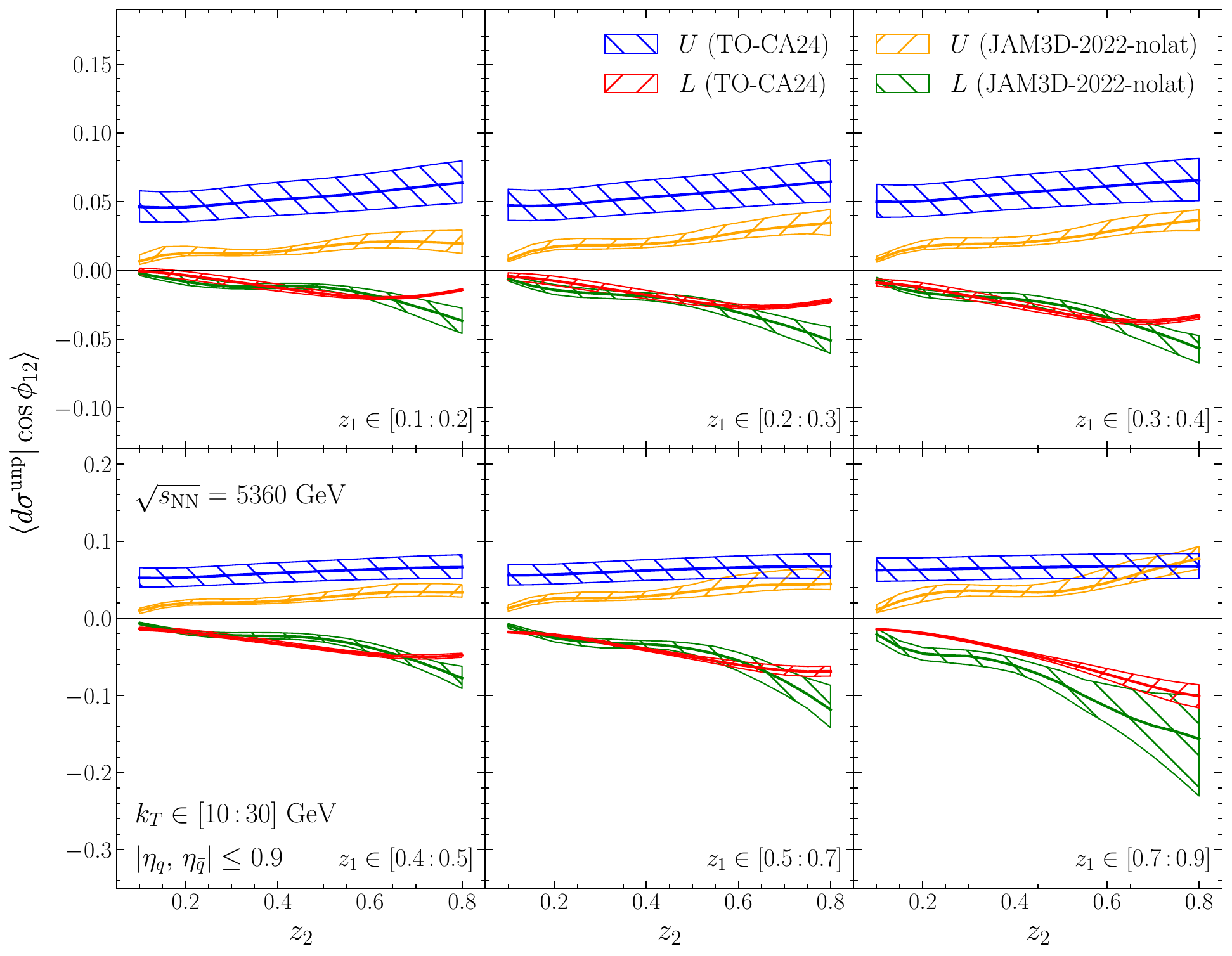}
}
\\
\subfloat[LHCb]{\label{fig:UPC-LHCb}\includegraphics[width=8.9cm, keepaspectratio]{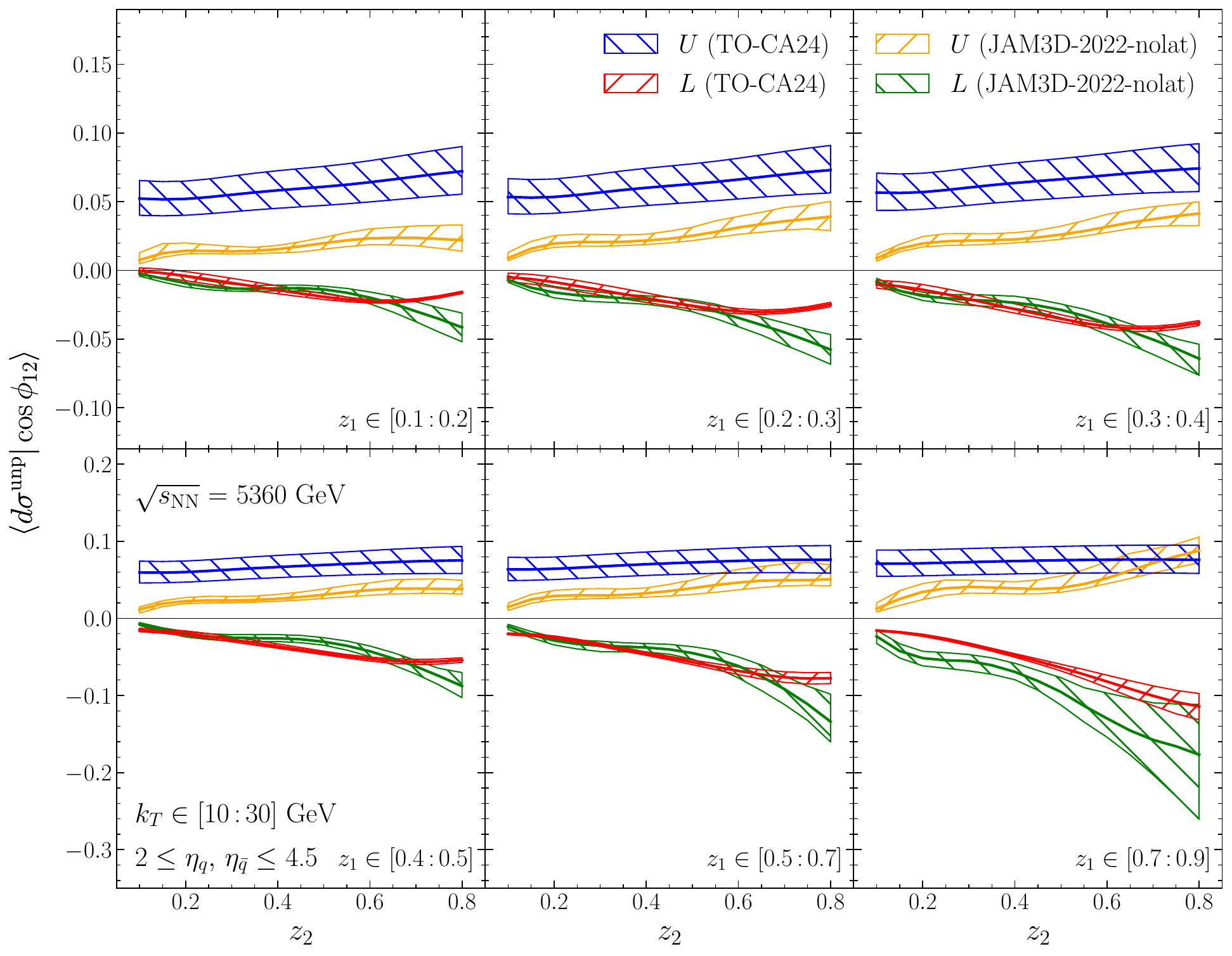}
}
\caption{The azimuthal moment $\langle {\rm d}\sigma^{\rm unp} \vert \cos\phi_{12} \rangle$ in PbPb UPCs at the LHC and $\sqrt{s_{\rm NN}} = 5.36$ TeV for the (a) ATLAS/CMS, (b) ALICE and (c) LHCb detectors. The color code is the same as in Fig.~\ref{fig:FCC-ee-asym}.}
\label{fig:UPC-HL-LHC-asym}
\end{figure}

At variance with FCC-ee, as already discussed in Section~\ref{sec:yy}, in the case of heavy-nuclei UPCs only unpolarized photons contribute. Therefore, in this case we present predictions only for the azimuthal moment $\langle {\rm d}\sigma^{\rm unp} \vert \cos\phi_{12}\rangle$ of the unpolarized cross section. Our predictions are gathered in Fig.~\ref{fig:UPC-HL-LHC-asym}, respectively for the ATLAS/CMS (Fig.~\ref{fig:UPC-CMS}), ALICE (Fig.~\ref{fig:UPC-ALICE}) and LHCb (Fig.~\ref{fig:UPC-LHCb}) detectors. Also in this case, we present predictions for the unlike-sign and like-sign combinations in Eq.~\eqref{eq:U-L-comb} and based on the \TOCA and \JAM3D extractions, using the same color code as in Fig.~\ref{fig:FCC-ee-asym}. 

As in the case of FCC-ee, the azimuthal moment has a different sign for the two combinations, being again positive and in general larger in size for the unlike-sign case, apart from the large $z_1$, $z_2$ domain where the like-sign combination leads to comparable and even larger absolute values.

Concerning the two extractions adopted, for all the three kinematical setups considered, in the unlike-sign case the \TOCA set gives predictions ranging approximately from 5\% at lower $z_2$ values up to 10\% in the large $z_2$ range. At the same time, in all cases the \JAM3D set leads to sizably lower predictions at the smallest $z_1$ bins, approaching the \TOCA results in the largest ones. In particular, the ALICE and LHCb cases in the lower $z_1$ bins could allow to distinguish between the predictions of the two extractions, if enough statistics can be collected. Let us stress however that the bands refer only to the statistical uncertainties and a full evaluation of the theoretical uncertainty is very difficult to reach and has not been performed yet.
As for the like-sign case, the two extractions give very similar results, ranging from a few percent up to 10-20\% in size at large $z_1$, $z_2$ values. The only significant differences can be seen in the largest $z_2$ bins, where however the uncertainty band for the \JAM3D extraction is larger and overlap with the one of the \TOCA case.

Finally, let us remark that most of these trends are qualitatively similar to those described in the FCC-ee case (see Fig.~\ref{fig:FCC-ee-asym} and the corresponding comments in the text).

We expect that these predictions may be tested already using the LHC Run 3 data, or in the future at the HL-LHC. Once again, such measurements could provide new data to improve the current phenomenological extractions and to disentangle among different models for the TMD quark fragmentation functions.

%% file: Conclusions.tex
\section{Conclusions and outlook}
\label{sec:conc}

In this paper we have investigated the Collins effect for the inclusive production of charged-pion pairs in the quasi-real photon-photon fusion channel  $\gamma_1 \gamma_2 \to q \bar q \to h_1 h_2 X$, both in  $e^+e^-$ SIA annihilation processes, $e^+e^- \to  h_1 h_2\, X$, and in UPCs, ${\cal N}_1  {\cal N}_2\to  h_1 h_2\, X$. Within the helicity formalism and assuming TMD factorization, we have calculated, at leading order in perturbative QCD, the azimuthal asymmetries in the distributions of the two almost back-to-back pions around the jet thrust-axis, arising from the Collins effect. Should these asymmetries be sufficiently large to be measured, comparing our predictions with future data will allow us to use these processes as tools to gather complementary information on the quark TMD fragmentation functions. This would enable one to probe different energy scales and flavor combinations of the fragmentation functions as compared to SIDIS and SIA processes, which are the reactions typically studied for this purpose and for which data are presently available.

We have performed our analysis by adopting two different parametrizations of the Collins and unpolarized FFs available in the literature~\cite{Gamberg:2022kdb,Boglione:2024dal}. We have then provided numerical estimates of the azimuthal asymmetries for  kinematic configurations accessible in $e^+e^-$ collisions at the planned FCC-ee~\cite{FCC:2025lpp} and in lead-lead UPCs at the LHC in its future High-Luminosity phase~\cite{ATLAS:2023dns,ATLAS:2024mvt,CMS:2023gfb,ATLAS:2025aav,CMS:2022lbi,CMS:2025lsm,ALICE:2023udb,LHCb:2022ahs,LHCb:2023hlw}. Such experiments differ in the values of $\sqrt{s}$ and, in the LHC case, also in the pseudorapidity ranges covered by the different detectors.
We have identified the regions of phase space where the azimuthal asymmetries can be sizable, even up to around 10-15\%. It turns out that there are some kinematic regions where the two parametrizations give different predictions for the asymmetries, making it possible, in principle, to discriminate  between the two models for the Collins function, provided that measurements in those regions have sufficient precision.

We therefore believe that our proposed measurements could contribute to improve the current phenomenological extractions of TMD quark fragmentation functions. As this also represents a first and exploratory work, it remains to be seen how stable our predictions are with respect to possible improvements of the theoretical framework adopted, such as, for instance, the implementation of the TMD evolution of the fragmentation functions and the inclusion of the transverse momentum dependence in the photon distributions, along the lines of the analyses in Refs.~\cite{Zhang:2022tee,Shi:2024gex,Boer:2024cnw} for dilepton, dijet and heavy-quark pair production in UPCs. Additionally, one can easily extend the formalism to the so-called hadron frame configuration~\cite{DAlesio:2021dcx}, in which one consider the azimuthal asymmetries of one of the hadron with respect to the direction of motion of the second hadron in the opposite hemisphere, without requiring the thrust-axis recostruction, that can be experimentally challenging. 
Moreover, future extractions of the fragmentation functions can be used to update our results.

In the future, we also plan to extend our studies to the case of almost back-to-back pion pair production in photon-photon fusion processes in electron-nucleus collisions at high-energies. Such experiments could be performed at the future Electron-Ion Collider planned in the US~\cite{AbdulKhalek:2021gbh}.
Finally, our study can be easily extended to the production of other mesons, like kaons, and to spin 1/2 barions, like $\Lambda$ particles, in the spirit of Refs.~\cite{DAlesio:2021dcx,DAlesio:2023ozw}.